\documentclass[structabstract]{aa}

\usepackage{amsmath,array,amsfonts,mathrsfs}

\usepackage{float}
\usepackage{graphicx}
\usepackage{amssymb}
\usepackage{txfonts}
\usepackage{natbib}

\begin{document}

\newcommand{\OHb}{[OIII]$\lambda5007$/$H_{\beta}$~}
\newcommand{\NHa}{[NII]$\lambda6584$/$H_{\alpha}$~}
\newcommand{\NII}{[NII]/$H_{\alpha}$~}
\newcommand{\xmm}{\emph{XMM-Newton }}
\newcommand{\OIII}{[OIII]$\lambda5007$~}
\newcommand{\lum}{erg~s$^{-1}$}
\newcommand{\flux}{erg~s$^{-1}$~cm$^{-2}$}
\newcommand{\Hb}{H$_{\beta}$}
\newcommand{\Ha}{H$_{\alpha}$}
\newcommand{\lx}{$L_X$}
\newcommand{\he}{$2-10$}
\newcommand{\se}{$0.5-2$}

\title{The X-ray luminous galaxies optically classified as star forming are mostly narrow line Seyfert 1s.}

   \author{N.~Castell\'o-Mor \inst{1}\fnmsep\thanks{\email{castello@ifca.unican.es}},
          X.~Barcons\inst{1},
          L.~Ballo\inst{1},
          F.J.~Carrera\inst{1},
	      M.J.~Ward\inst{2},
          C. Jin\inst{2}
          }

   \institute{Instituto de F\'isica de Cantabria (CSIC-Universidad de Cantabria),
              E-39005 Santander, Spain
	\and
	      Department of Physics, 
	      University of Durham,
	      South Road,
	      Durham, DH1 3LE, UK
             }

   \date{Received 2011 October 20; accepted 2012, April 20}

   \abstract
   {The optical and ultraviolet emission lines of galaxies are widely used to distinguish star-forming galaxies 
   (SF) from active galactic nuclei (AGNs). However, this type of diagnostic has some associated uncertainties, 
   because AGNs can be of low luminosity and/or heavily obscured, and the optical emission lines 
   may be dominated by a stellar component. On the other hand, and despite its limitations,
   X-ray emission can be used as a reliable tracer of luminous AGNs. Several well-studied 
   examples exist where the optical diagnostics are indicative of SF galaxy, but the X-ray properties reveal 
   the presence of an AGN.}
   {We aim to characterize the nature of galaxies whose optical emission line diagnostics 
   are consistent with star formation, but whose X-ray properties strongly point towards the presence of 
   an AGN. Understanding these sources is of particular importance in assessing the completeness of AGN samples 
   derived from large galaxy surveys, selected 	solely on the basis of their optical spectral properties.}
   {We construct a large sample of $211$ narrow emission line galaxies (NELGs) (which have full widths at half
   maximum (FWHMs) \Hb~ emission line $<1200$ km/s) from the SDSS-DR7 galaxy spectroscopic catalogue, 
   for which we are able to construct a classical diagnostic diagram, 
   [OIII]/\Hb~ versus [NII]/\Ha~ (hence $z<0.4$), and that are also detected in the $2-10$ keV 
   X-ray band and present in the 2XMM X-ray source catalogue. This sample offers a large database by 
   which to investigate potential mismatches between optical diagnostics and X-ray emission.}
   {Among these $211$ objects, which based on our selection criteria all are all at $z<0.4$, we find that 
   $145$ galaxies are diagnosed as AGNs, having $2-10$ keV X-ray luminosities that span a wide range, 
   from $10^{40}$ erg/s to above $10^{44}$ erg/s. Out of the remaining $66$ galaxies, which are instead 
   diagnosed as ``star-forming'', we find a bimodal distribution in which $28$ have X-ray luminosities in excess 
   of $10^{42}$ erg/s, large thickness parameters ($T =  F_{2-10 keV}/F_{[OIII]} 
   > 1$) and large X-ray to optical flux ratios ($X/O>0.1$), while the rest are consistent with being simply starforming
   galaxies. Those $28$ galaxies exhibit the broadest $H_{\beta}$ line widths (FWHMs from $\sim300$ to $1200$ 
   km/s), and their X-ray spectrum is steeper than average and often displays a soft excess. }
   {We therefore conclude that \emph{the population of X-ray luminous NELGs with optical lines 
   consistent with those of a starforming galaxy (which represent $19$\% of our whole sample) is largely dominated by 
   narrow line Seyfert 1s (NLS1s)}. The occurrence of such sources in the overall optically selected sample is small
   ($<2\%$), hence the contamination of optically selected galaxies by NLS1s is very small.}


   \authorrunning{Castell\'o-Mor et al.}
   \titlerunning{X-ray luminous, optically SF galaxies}
   \maketitle
%


\section{Introduction}
\label{section1}

Emission lines in galaxies convey information about their emitter such as the power and nature of the
underlying exciting source, as well as the geometry, physical condition and chemical composition of the gas, among other
properties. Emission line data is available for many galaxies, which allow them to be classified 
depending on their physical nature, as either star forming (SF) or hosts of active galactic nuclei (AGNs). 
On the other hand, determining the fraction of galaxies hosting AGNs at their centres is an essential step for 
studies of galaxy evolution. The way in which AGNs are usually identified among the large samples of galaxies 
in massive optical spectroscopic surveys is almost invariably in terms of their emission lines. 
\cite{Baldwin1981} were amongst the first to introduce robust emission-line diagnostic ratios able to
distinguish between star-forming processes and active nuclear emission. These and other diagrams were later used by 
\cite{Veilleux1987} to derive a semi-empirical classification scheme to distinguish between AGNs and SF galaxies.
\cite{Kewley2001} used these same diagrams to derive a purely theoretical classification scheme which was
later extended by both \cite{Kauffmann2003} and \cite{Stasinska2006}. The underlying idea is that the emission 
lines in normal SF galaxies are powered by massive stars, so there is an upper limit to the intensity ratio of the 
collisionally excited lines with respect to the recombination lines (such as \Ha~or \Hb). 
In contrast, the photons from an AGN extend to yet higher energies and therefore induce more heating, implying 
that optical collisionally excited lines should be brighter with respect to recombination lines than in the case 
of ionization only by massive stars. \\

However, as noted by a number of authors \citep[e.g., ][]{Severgnini2003,Page2006,Caccianiga2007,Trump2009},
emission lines can be hidden, diluted, or masked by the stellar light from the galaxy. This is a particular problem
in a low luminosity AGN (LLAGN), where the observed star formation and AGN components may be of comparable brightnesses. 
In these cases, the evidence of AGN activity in optical spectra can be weakened in a significant number of objects. 
In other cases, the regions producing the narrow and broad line emission characteristic of an AGN, 
may be obscured by dust in the host galaxy and/or in the nuclear regions 
\citep[see][]{Iwasawa1993,Comastri2002,Rigby2006,Civano2007}. Such objects would be optically classified 
as either SF galaxies or HII regions, on the basis of emission-line diagnostic diagrams. Therefore, emission
line diagnostics are not always a reliable means of detecting AGNs within galaxy samples. \\

As an alternative tracer, X-ray emission is a virtually universal feature of the AGN phenomenon. Accretion onto the super
massive black hole (SMBH) produces X-ray emission by means of inverse Compton scattering by a hot electron corona of 
ultraviolet (UV) photons emitted from the accretion disc. Hard X-rays are not strongly attenuated along their path 
from the central engine, except by extremely high gas columns of $\gtrsim 2\times10^{24}$ cm$^{-2}$, which occur in 
the so-called Compton thick sources. The latter are believed to make a significant contribution to the infrared (IR) 
and submillimeter backgrounds, because of the absorption of a large fraction of their continuum at shorter wavelengths, 
and its reradiation at longer wavelengths.
This gas column density is equivalent to several tens of magnitudes of optical extinction, 
for Galactic gas to dust ratios. Contemporary X-ray observatories such as NASA's {\it Chandra} and ESA's 
\xmm are sensitive to photon energies of up to $\sim10$ keV, thus detection by these facilities is a very 
robust indicator of AGN activity. 
However, in spite of this advantage of X-ray selection there is no single method capable of selecting
a complete sample of AGNs.  On the one hand, dust is the main problem for optical selection, particularly for
edge-on disc galaxies. On the other hand, X-ray selection is biased against sources in which the X-ray emission 
is heavily absorbed and/or Compton-scattered by dense gas clouds close to the central engine. 
Thus, no single method is able to identify all the AGNs found by other methods. \\

When both X-ray data and optical spectra are available for the same galaxy, galaxies optically classified as SF, 
may be found to have high X-ray luminosities, in excess of the most luminous SF 
galaxies known in the local Universe, by over an order of magnitude ($>10^{42}$~erg~s$^{-1}$). 
The origin of this classification discrepancy is not fully
understood. \citet{Trouille2010} compared the optical classifications with the X-ray properties of a complete sample
selected from three {\it Chandra} fields. They found that the optical diagnostic diagram misidentified $20-50\%$ of their
X-ray selected AGNs, in the sense that many X-ray AGNs were misclassified as SF galaxies.
\citet{Yan2011} reported another case of misclassification. Their analysis of the relationship between 
the X-ray properties and optical emission lines suggests that a large fraction of the X-ray emitting AGNs
would not have been identified using emission line diagnostic diagrams. They also found that there are indications of 
large classification discrepancies between X-ray and optically selected AGN samples at $z\sim1$.\\

Understanding the physical cause of why a fraction of galaxies exhibit emission line diagnostics compatible with
star formation, yet have X-ray properties that are indicative of an AGN, is of considerable importance. It must relate 
to the demographics of AGNs, black hole growth, and galaxy evolution. To address these issues, we present a study 
of a large sample of narrow emission line galaxies (NELGs) from the Sloan Digital Sky Survey (SDSS), 
whose X-ray emission properties are all available from the 2XMM X-ray source catalogue. We devote special attention
to those objects optically classified as SF galaxies, but with $2-10$ keV luminosities in excess of $10^{42}$ erg/s.
We conduct a detailed analysis of this population, using other optical spectral features, e.g. the full width at half
maximum (FWHM) of the \Hb~ emission line, as well as the X-ray spectral properties. We discover that this optically 
misclassified population, which represents over 10\% of our full sample of narrow emission line galaxies with detected 
X-ray emission, consists of narrow line Seyfert 1 galaxies (NLS1s).
If, however, all narrow emission-line galaxies for which there is \xmm X-ray coverage (regardless of whether they have
an X-ray detection or only an upper limit) were considered, the fraction of misidentifications would be only $1.5-5 \%$.
\\

The structure of the paper is as follows. In Section \ref{section2}, we select the sample and describe both  
classification methods, highlighting the disagreements between them. Our modelling of the X-ray data is presented in 
Section \ref{section3}, which focusses on the general spectral features found for the whole sample.
Finally, in Section \ref{section4} we discuss our results and their implications. 
We assume a concurrence cosmology with $H_0=70$ \rm{km s}$^{-1}$ \rm{Mpc}$^{-1}$, $\Omega_{\Lambda}=0.73$, 
and $\Omega_M=0.27$. 

\section{Optical and X-ray properties}
\label{section2}

To build a sample of galaxies having both X-ray data and optical spectra and showing prominent narrow emission lines 
and a lack of broad components, we performed a cross-correlation between the 2XMMi catalogue and the spectroscopic SDSS DR7 
catalogue\footnote{The SDSS DR7 data archive server is available at http://www.sdss.org/dr7}. The SDSS DR7 is the seventh 
major data release of the Sloan Sky Survey. The SDSS spectroscopic data has sky coverage of $\sim 8200$ deg$^2$, a spectral 
coverage from $3800$ \AA{} to $9200$ \AA{}, and a spectral resolution ranging from $1850$ to $2200$.

We filtered the resulting large sample as described below. Our final sample is composed of 211 NELGs, with \Hb~
line widths ranging from $140$ km~s$^{-1}$ up to $1200$ km~s$^{-1}$.  The X-ray selection criteria resulted in all spectra 
having a minimum of 30 counts above $2$ keV in at least one detector, ie. PN, MOS1, or MOS2.

\subsection{Sample selection}

Our sample selection process consisted of five stages:
\begin{enumerate}
 \item \emph{Identification of NELGs}. The first step was to obtain a large sample of NELGs from the SDSS DR7 catalogue. 
 We adopted an operational line width cut-off FWHM(\Hb)$\leq 1200$ km~s$^{-1}$, to reject galaxies with broad emission lines. 
 This value was chosen by taking into account the strongly bimodal distribution of measured \Hb~ FWHM values for an 
 emission-line galaxy sample \cite[see][Fig.~6]{Hao2005}. This indicates that there is a natural separation between broad 
 and narrow line AGNs. Our selection results in a sample composed of objects of different types including: type 2 AGNs with 
 a broad range of luminosity, obscured AGNs, galaxies whose emission is not dominated by nuclear activity and are classified 
 as normal star-forming galaxies, and type 1.9 Seyferts, which exhibit only a weak broad component of \Ha~ \citep{Osterbrock1981}.\\
 \item\emph{BPT diagram as indicator of SF/AGN activity}. The standard method used to classify galaxies as either SF or AGNs, 
 is based on their emission-line ratios. Among all available combinations of emission lines, in this  work we used \OHb 
 versus \NHa \citep[often referred to as the BPT diagram]{Baldwin1981}, because it is the one that most clearly distinguishes 
 between these two classes \citep{Stasinska2006}. In the BPT diagram (see Figure~\ref{fig:BPTdiagram}), the AGNs 
 occupy the region above and to the right of the borderline,
 whilst SF galaxies are found to the bottom left of the parameter space  \citep{Kewley2001, Kauffmann2003}. The
 reason for using these emission-line ratios is twofold: first, the lines  used to compute each ratio are very 
 close together in wavelength,  and consequently the line ratios are largely insensitive to dust; 
 second, the \OIII emission line luminosity is often used as an isotropic indicator of AGN activity.  
 This assumes that the [OIII] emission is an unbiased and
 orientation-independent measure of the ionizing flux from the AGNs whereas X-ray photons from the compact nucleus may be
 strongly attenuated in the plane of the dusty torus. Therefore we only selected objects in which the four emission lines 
 of \Hb, \Ha, \OIII, and [NII]$\lambda6584$ were detected. On the basis of this selection criteria all of them are at 
 redshift $z<0.4$. Given the typical signal-to-noise (S/N) ratio and the instrumental resolution of the SDSS spectra, we omitted 
 objects that had an observed FWHM for any of the four emission lines smaller than $70$ km~s$^{-1}$, as they are likely to be 
 spurious detections or poorly detected lines. Applying these criteria, we selected about 150000 nearby NELGs.
 \item\emph{Cross-correlation between 2XMMi DR3 \& SDSS  DR7}. To identify possible X-ray counterparts, 
     we analysed \xmm observations covering the sky positions of these NELGs. A total of 1729 have some X-ray exposure time,
     although in the majority of cases only upper limits to their X-ray emissivity were found. We crosscorrelated
     this parent sample with the Incremental Second XMM-Newton Serendipitous Source Catalogue \emph{2XMMi-DR3} \citep{Watson2009}
     released in April 2010\footnote{Available from http://xmm.esac.esa.int} applying a matching radius of 3 arcsec.
     This radius is chosen as a compromise between allowing for some positional error, and minimizing the probability of spurious
     matches \citep{Watson2009}. We considered only X-ray  sources with a $0.2-12$ keV European Photon Imaging Cameras EPIC 
     detection likelihood above $3\sigma$ in at least one camera.
 \item \emph{\lx~as an indicator of AGN activity}. Our primary goal is to characterize the populations of NELGs by comparing 
     their optical classification with their X-ray properties. The standard technique employed to identify AGNs in emission 
     line galaxies is via an empirical X-ray luminosity threshold at $L_X>10^{42}$ \lum. This is a very conservative limit, 
     based on the fact that very few local starforming have higher X-ray luminosities -with a few possible exceptions e.g. 
     NGC3256 \citep{Lira2002}. While very luminous AGNs can be unambiguously identified in almost any energy band, 
     AGNs become progressively more challenging to identify at lower luminosities when their emission may be equal to, 
     or even less than that from the host galaxy. The hard X-ray (\he~keV) luminosity (\lx) is a good indicator of AGN activity, 
     because the X-ray spectra of SF galaxies are typically softer than those of AGNs. However, lower luminosity X-ray sources 
     with hard spectra, can be either AGNs or arise from high mass X-ray binaries. Hence there is a problem in using X-rays alone
     to distinguish between low luminosity or obscured AGNs, and a population of high mass X-ray binaries in 
     starforming galaxies. One way to distinguish between these possibilities is to apply the empirical limit
     in X-ray luminosity of starburst galaxies at around $L_X \sim 10^{41}$ \lum. Therefore, a factor of 
     ten above this value, it is assumed that the X-rays probably arise from accretion onto a SMBH.
     We have adopted this value as  the representative dividing line for starforming activity: 
     all objects with an X-ray luminosity higher than $10^{42}$ \lum~ are assumed to host an active nucleus. Galaxies with 
     lower X-ray luminosities are consistent with SF galaxies. But it is possible that low luminosity X-ray sources may still
     be weak AGNs, e.g. low-ionization nuclear emission-line regions (LINERs), or be are heavily obscured. 
     Other information, such as X-ray spatial extent or variability, is needed to confirm the origin of the emission.
     To select galaxies that might potentially host an AGN, we required our sources to have a well-defined count rate in 
     the $2-12$ keV energy range, i.e. we considered only X-ray sources with a $2-12$ keV European Photon Imaging Cameras (EPIC) 
     detection likelihood above $3\sigma$ in at least one camera. This requirement resulted in X-ray spectra with a minimum 
     of $30$ counts in at least one detector.
     \item \emph{Final sample}. Finally, we performed a visual inspection of the optical data in each of these SDSS/2XMMi 
     pairs to confirm that all the matches were indeed genuine. Special care was taken to examine sources that showed some 
     signs of \Ha~and/or \Hb~ broad emission lines, as well as spurious sources. After inspection of the SDSS spectra, we excluded 
     $29$ objects. These sources showed either strong reddening or low S/N in the blueward part of the \Hb~ line or only weak 
     broad \Ha~ lines (e.g. were Seyfert 1.9's). In some cases, we could detect weak broad \Ha~ and \Hb~ lines (e.g.a Seyfert 1.8). 
     After removing these sources, our final sample contained $211$ galaxies, that had only narrow emission lines and reliable X-ray
     flux detections in the $2-12$ keV band.
\end{enumerate}

\subsection{Optical classification versus X-ray emission}

The BPT diagrams have been used to infer whether the gas in a given galaxy is excited by star formation or radiation from
an accretion disc around a central SMBH and has become one of the major tools for the classification and analysis of emission
line galaxies in the SDSS \citep{York2000}. To be conservative in our analysis, we adopted the dividing line between SF
and AGN galaxies presented by \citet[hereafter Kauf03]{Kauffmann2003}. In that work, a refined optical classification was
obtained, based on a combination of stellar population synthesis models plus detailed self-consistent photoionization models, in
order to create a theoretical starburst line projected onto the BPT diagram. This is given by

\begin{equation}
\log \text{\OHb} = \frac{0.61}{\log{( \text{\NHa} )-0.05}} + 1.3.
\end{equation}

\citet{Kewley2001} used a different separation criterion between SF and AGN galaxies in the BPT diagram, which lies
above and to the right of the Kauf03 line, defining a region often known as the \emph{LINER/transition region}
where sources exhibit both AGN and starburst activity. If we adopted the Kewley et al.
(2001) borderline, then the fraction of X-ray luminous NELGs classified as SF galaxies would be significantly higher, 
at least twice \citep[see e.g.][]{Jackson2012}, where their three sources fall between both borderlines). In this paper, 
we prefer to focus on the X-ray luminous NELGs that are uncontroversially classified as SF galaxies using the BPT diagram, 
hence we adopt the Kauf03 criterion. Following this criterion, we can directly obtain an optical classification for our sources:
star-forming galaxies are those lying below the Kauf03's line, i.e. \emph{optically classified} SF or BPT-SF, 
and conversely those located above the line are classified as AGN, i.e. as either an \emph{optically classified} AGN 
or BPT-AGN.\\

\begin{figure*}[!Hht]
\centering
        \includegraphics[width=0.7\textwidth]{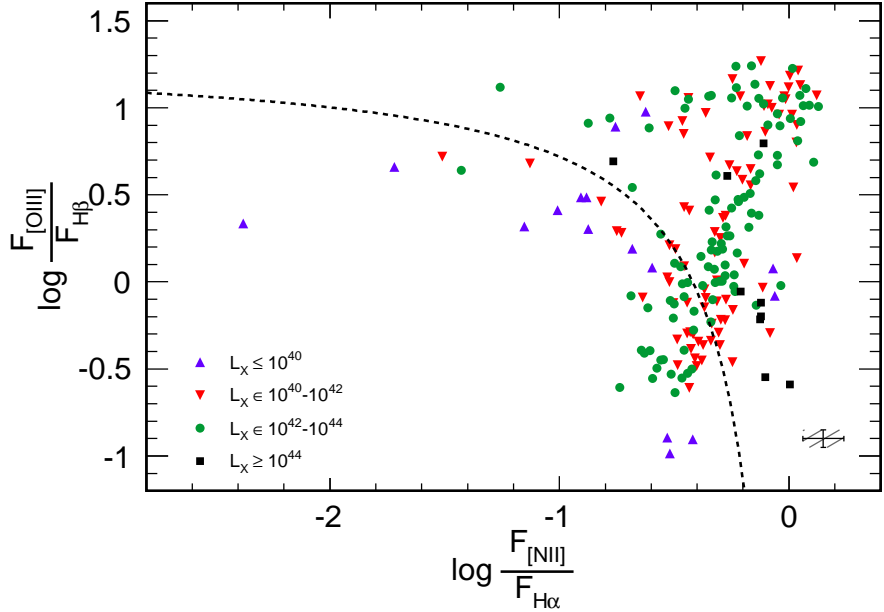}
        \caption{Emission line diagnostic diagram (BPT diagram) for the whole sample of NELGs. The curve separating the AGNs from 
        the non-AGNs (bottom left) zone is taken from \citet{Kauffmann2003}. Symbols change in both form and colour according to 
	the hard X-ray luminosity, \lx. To avoid confusion, only the mean errors are reported (bottom right). 
	The estimated errors increase with decreasing $log[NII]/$\Ha.}
        \label{fig:BPTdiagram}  
\end{figure*}

To see how the optical classification compares with X-ray luminosities, we must first obtain the hard X-ray intrinsic
luminosity of each source. We used the X-ray fluxes and spectroscopic redshifts\footnote{Optical and X-ray parameters are taken 
from SDSS-DR7 and 2XMMi-DR3, respectively.} to calculate the rest-frame \he~keV luminosities (\lx) assuming an X-ray spectrum 
in the form of a power law modified by Galactic absorption with continuum spectral slope $\Gamma=1.7$; we then corrected 
these luminosities for the Galactic absorption to compute the intrinsic luminosities, but did not correct for any possible 
intrinsic absorption. Figure~\ref{fig:BPTdiagram} shows the BPT diagram where the dashed line is the Kauf03 criterion and 
the points change in both form and colour according to their value of \lx~to highlight the disagreements between the 
optical-based and X-ray-based classifications. From the optical diagram and according to the values of \lx~the sample 
was split into four subsamples (see Table \ref{tab:subsamples}):

\begin{itemize}
 \item {\bf Weak-AGN subsample}: consisting of $62$ sources classified as AGNs according to the BPT diagram 
 notwithstanding their low luminosities, not exceeding $10^{42}$ \lum.
 \item {\bf Strong-AGN subsample}: including $83$ NELGs identified as BPT-AGNs that have in turn luminosities
 exceeding $10^{42}$ \lum.
 \item  {\bf True-SF subsample}: consisting of $38$ sources that were classified as SF according to both the BPT 
 diagram and \lx~ criterion.
 \item {\bf Missing-AGN subsample}: involves $28$ objects classified as BPT-SF that notwithstanding have a 
 \lx$\geq 10^{42}$ \lum.
\end{itemize}

\begin{table*}[!Hht]
    \caption[]{Classification definitions of NELG objects.}
    \label{tab:subsamples}
    \centering
    \begin{tabular}{|rl|c|c|c|c|c|c|}\hline
        \multicolumn{2}{|c|}{Sub-sample}& Optical$^{1}$ & X-ray$^{1}$ & N & T$\geq 10$ & X/O$\geq0.1$ & FWHM$_{H_{\beta}}>600$km/s \\
            &          &       &       &   & \%    & \% & \% \\ \hline
        \emph{BPT-SF} & \emph{true-SF}     & SF & SF/AGN  & 38 & 0.10 & 0.00 & 0.03 \\
                 & \emph{missing-AGN} & SF & AGN & 28 & 0.93 & 0.96 & 0.89  \\ \hline
        \emph{BPT-AGN}& \emph{weak-AGN}   & AGN & SF/AGN & 62 & 0.03 & 0.08 & 0.05   \\  
                 & \emph{strong-AGN}    & AGN & AGN & 83 & 0.30 & 0.79 & 0.19 \\\hline
    \end{tabular}
    \tablefoot{ \emph{Col. 1}: sub-sample name; \emph{col.2}: classification of the source using the Kauf03's line as a 
    separation criterion between SF and AGN activity; \emph{col. 3}: X-ray based classification according to hard X-ray 
    luminosity, \lx; \emph{col. 4}: total number of sources in the subsample; \emph{col.4(5)}: the fraction of sources in 
    the subsample that display a thickness parameter (X-ray-to-optical flux ratio) higher than $1$ ($0.1$) as expected 
    for a typical AGN; \emph{col. 5}: the fraction of sources that exhibit an \Hb FWHM larger than 600 km/s.}
\end{table*}

This classification clearly implies that there is a mismatch between the optical-based and X-ray-based classifications 
in the missing-AGN and weak-AGN subsamples \citep[also found by][]{Yan2011}.
On the one hand, there are several explanations of the low luminosity emitted by weak-AGNs.
A significant fraction of the population of AGNs in the local Universe displays a low X-ray luminosity, not
exceeding  $10^{42}$ \lum. \citep[see][LLAGN]{Barth2002}. In particular, low-ionization nuclear emission-line regions (LINERs) were
originally defined  by \citet{Heckman1980} as a subclass of these LLAGNs, whose optical spectra are dominated by strong 
low ionization lines and much weaker higher ionization lines classical AGNs. According to \citet{Heckman1980},
LINERs are galaxies that satisfy [OII]$\lambda3727 >$ [OIII]$\lambda5007$ and [OI]$\lambda6300/$[OIII]$\lambda5007 \geq 1/3$.
According to these criteria, we classified 8 (13\%) sources in our weak-AGN subsample as pure LINERs and an additional
18 (29\%) as weak-[OI] LINERs. The latter fully satisfy \citet{Filippenko1992}'s definition (i.e. 
[OII]$\lambda6300/$\Ha $<1/6)$. On the other hand, the high values of the luminosity emitted by the sources in our missing-AGN
subsample suggest that these sources do contain AGNs even though they lie beneath Kauf03's line implying that optical
AGN signatures are lacking and signs of star formation, such as strong \Ha~ and \Hb~ lines, are clearly visible.
The nature of this \emph{misclassified} population is discussed throughout this paper.

\subsection{Optical versus X-ray properties} 

After discussing the optical classifications of our NELG sample and the mismatches with the X-ray luminosities for some
objects, we now compare their optical and X-ray properties in the context of three parameters in an attempt to provide
clues about the nature of the source populations within the complete sample of NELGs.\\

\begin{figure*}[!Hhtb] 
        \centering
        \includegraphics[width=0.496\textwidth]{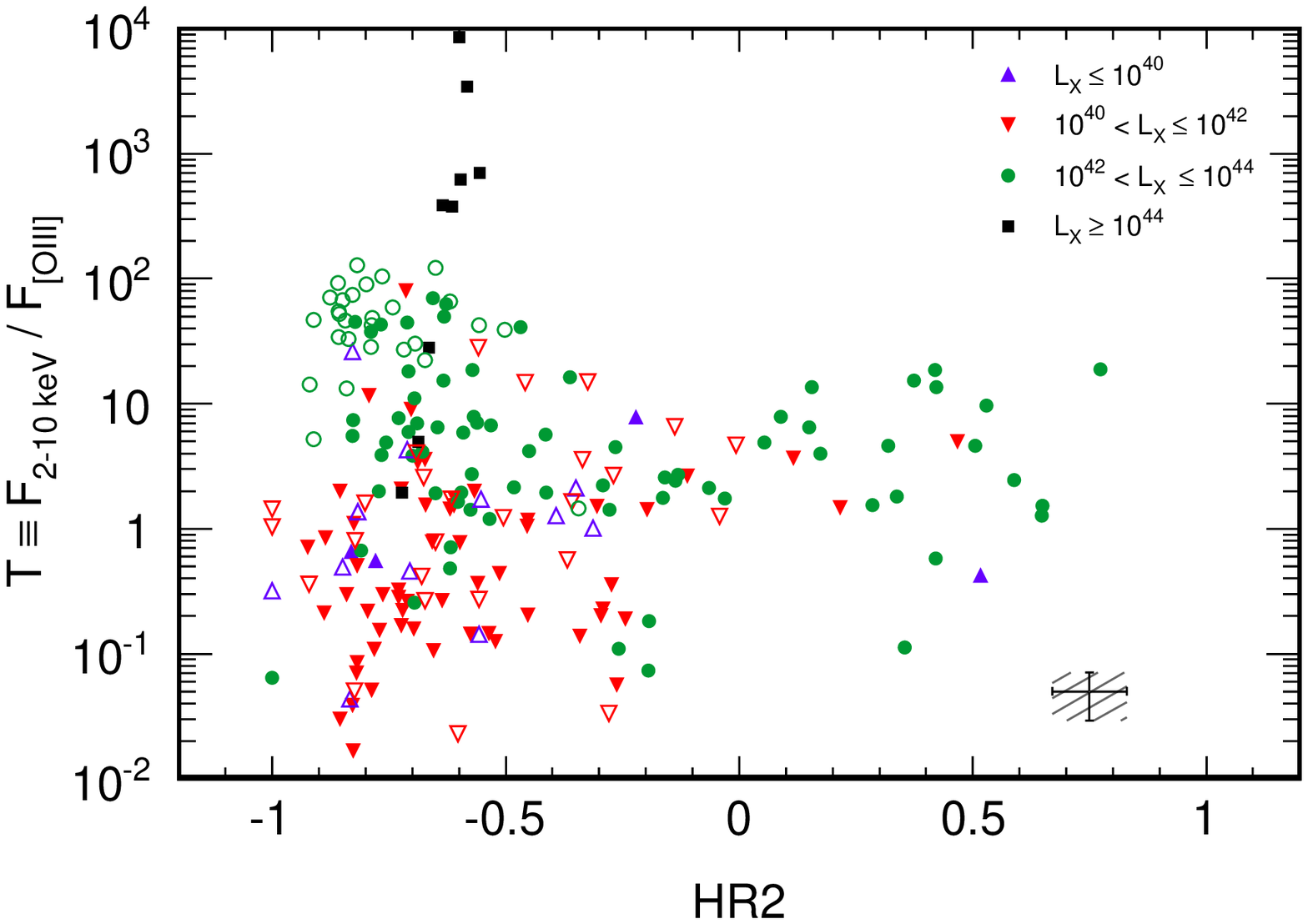}
        \includegraphics[width=0.496\textwidth]{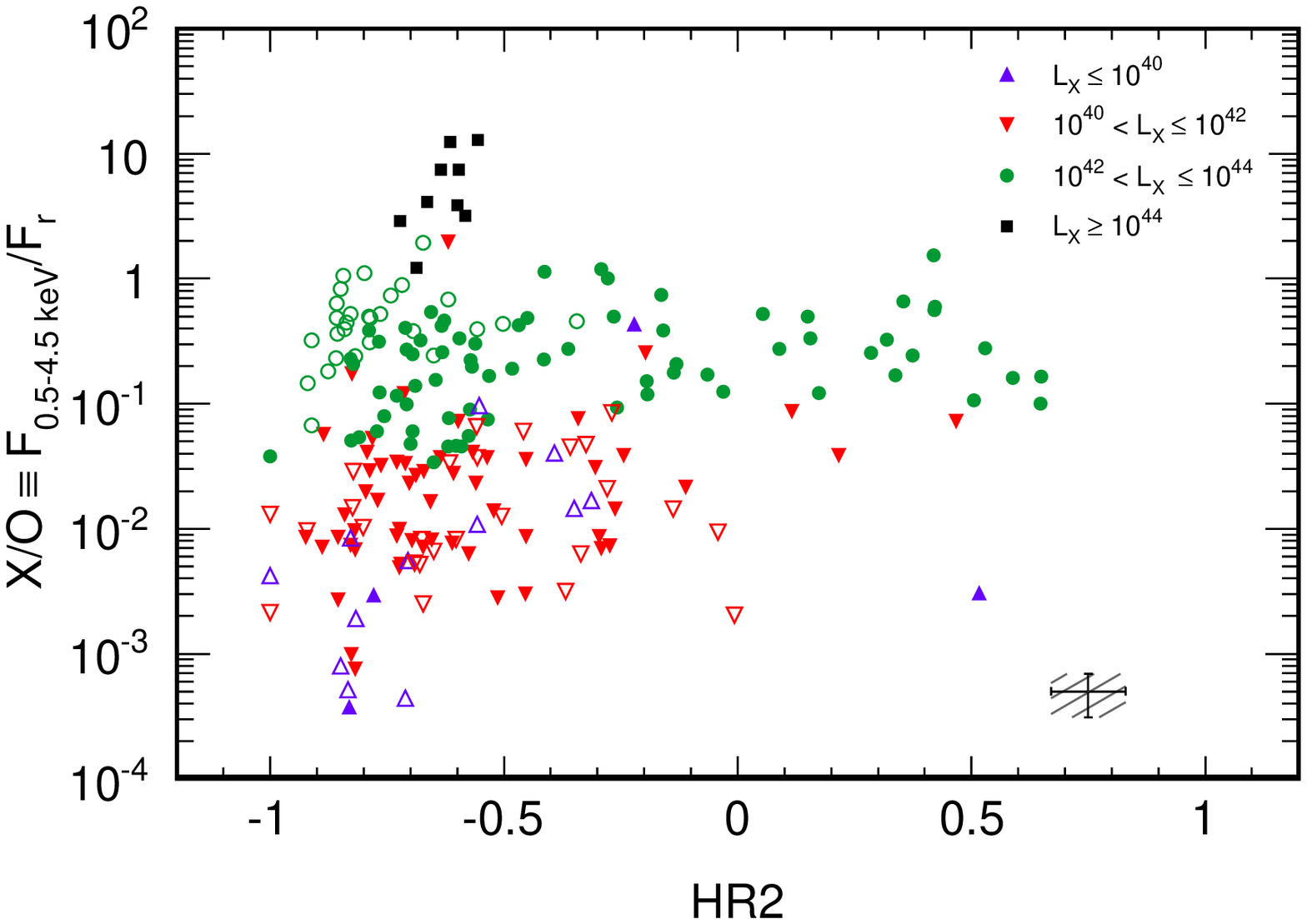}
        \includegraphics[width=0.48\textwidth]{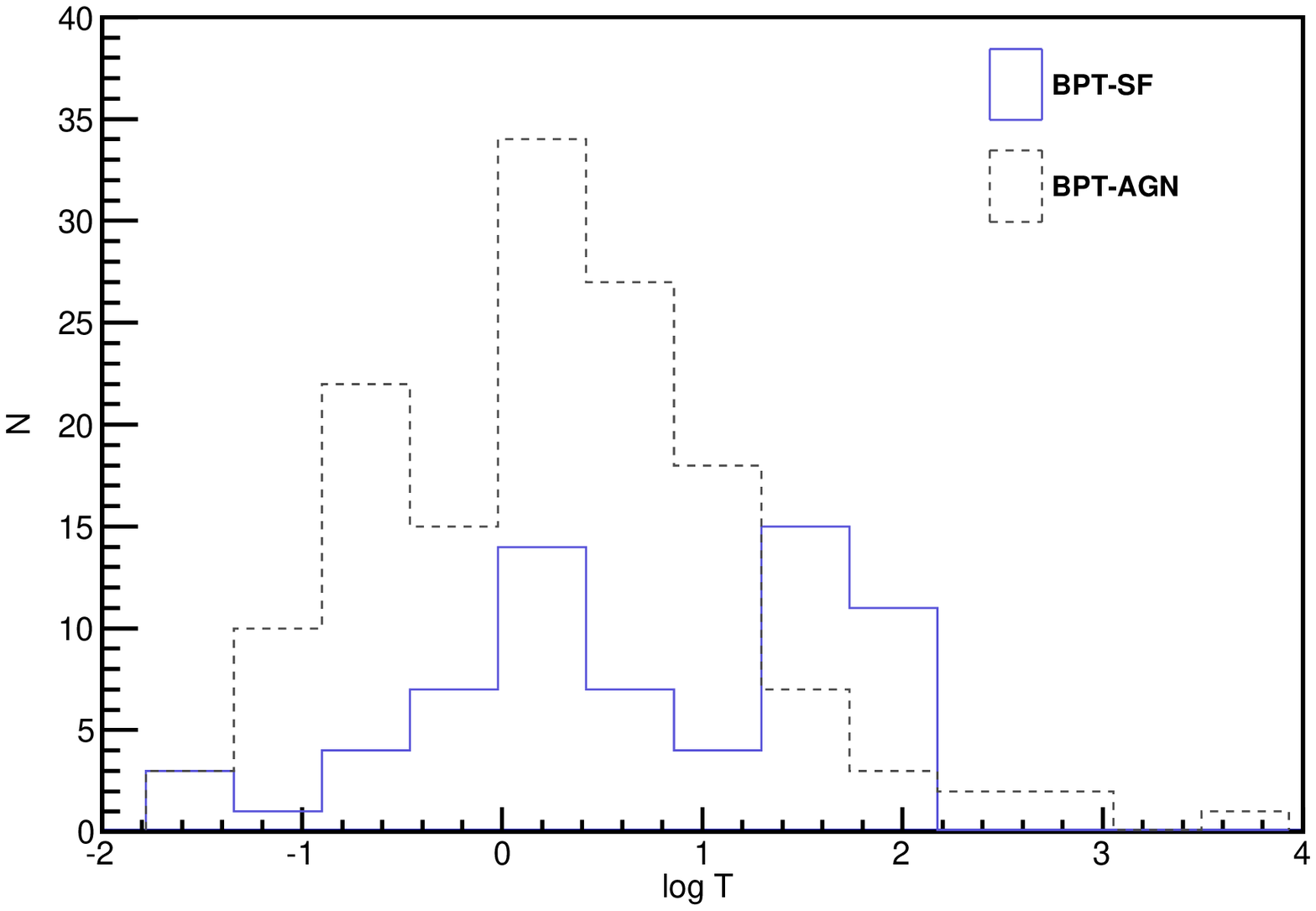}
        \includegraphics[width=0.48\textwidth]{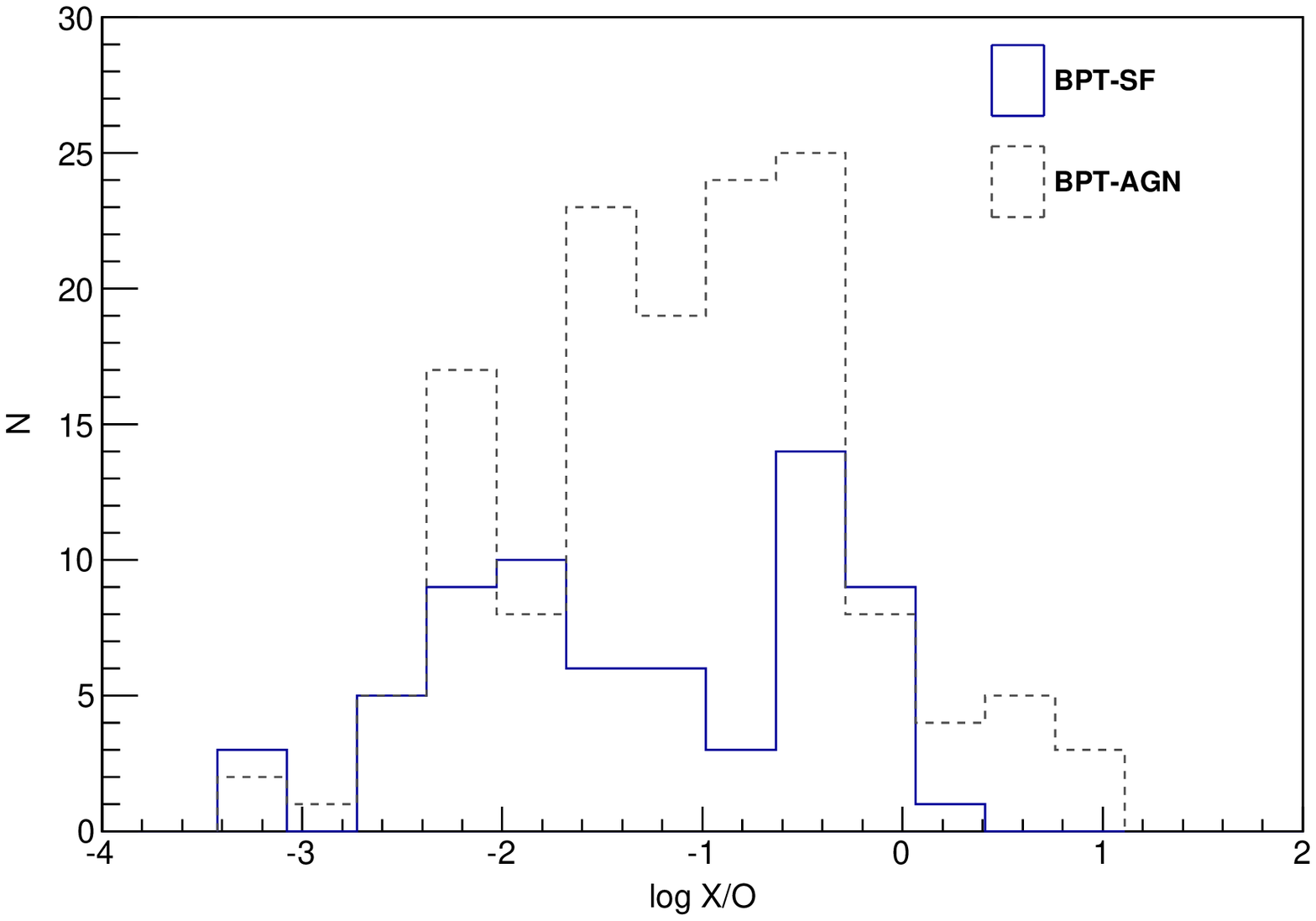}
	    \caption{ Optical and X-ray properties. \emph{Upper two panels:} $T = F^c_{HX}/F^c_{[OIII]}$ 
        versus $HR$ (\emph{on the left}) and the X-ray-to-optical flux ratio distribution as a function of HR 
        (\emph{on the right}). 
        Different symbols mark X-ray luminosity and the filled/empty symbols represent the optical classification 
        (BPT-AGN and BPT-SF, respectively). In both plots, the point with the error 
        bars is a fake source to represent the error mean value of the parameters. 
        \emph{Bottom two panels:} Distribution of the thickness parameter (\emph{on the left}) and the $X/O$ distribution
        (\emph{on the right}) for the two optical subsamples.}
        \label{fig:HR2_vs_T_and_XO}
\end{figure*}

To obtain the most basic X-ray spectral information, we performed a \emph{Hardness Ratio} (HR) analysis using EPIC-pn
data. We adopted the standard hardness ratio, defined as

\begin{equation}
        HR = \frac{H - S}{H + S},
\end{equation} 

\noindent
where S and H are the PSF and vignetting-corrected count rates in the \se~ keV and $2-4.5$ keV energy bands, respectively. A
HR analysis is much simpler than a complete spectral analysis and is often the only X-ray spectral information available
for the faint sources in the \xmm catalogue. We note that the X-ray selection criteria resulted in a minimum count
threshold of around $30$, hence a proper X-ray spectral analysis could not be performed for a number of our sources.
The HR parameter is an approximate indicator of the intrinsic X-ray spectral shape, which is also sensitive to the level 
of absorption. An unabsorbed X-ray spectrum has typically a lower HR than an absorbed one. 
Although this correlation is relatively weak, and redshift-dependent \citep{Trouille2009}, the vast majority of our missing-AGNs 
have a low HR that is consistent with being unabsorbed as shown in Figure~\ref{fig:HR2_vs_T_and_XO}.\\

An alternative method for evaluating absorption is to measure the X-ray luminosity, and compare this with an isotropic
indicator of the intrinsic power of the AGN. Assuming that the unified AGN model is correct, in absorbed sources the X-ray flux
iis attenuated with respect to this isotropic indicator by an amount related to the absorbing column density. Taking the 
reddening corrected \OIII luminosity as an isotropic indicator of the source nuclear strength, we calculated the ratio of the
hard X-ray to [OIII] fluxes \citep[hereafter thickness parameter or T]{Bassani1999}. According to \citet{Bassani1999},
Seyfert 1 galaxies lie in the range $1<T<100$, Compton-thin Seyfert 2 galaxies, in the range $0.1<T<10$, and Compton-thick Seyfert
2 galaxies at $T< 0.1$. To estimate the thickness parameter, $T\equiv F^c_{HX}/F^c_{[OIII]}$, X-ray and [OIII] fluxes 
were corrected for Galactic absorption and reddening, respectively. We used the \citet{Bassani1999} relation to derive 
$F^c_{[OIII]}$ from the Balmer decrement, as $F^c_{[OIII]} = F_{[OIII]} \left( \frac{(H_{\alpha}/H_{\beta})_{obs}}{3.0}
\right)^{2.94}$ which assumes an intrinsic Balmer decrement equal to $3.0$ as predicted in the NLR \citep[see][]{Osterbrock2006}.\\

Finally, a useful parameter used to discriminate between different classes of X-ray sources is the X-ray-to-optical flux
ratio \citep[see][hereafter $X/O$]{Maccacaro1988}. In this paper, we defined the X-ray to optical flux ratio using the observed
X-ray flux in the $0.5-4.5$ keV energy range and the optical $r(SDSS)$ band flux 
\citep[for the appropriate conversion factors see][]{Fukugita1995}. X-ray selected AGNs (of both type 1 and type 2) 
have typical $X/O$ flux ratios in the range between $0.1$ and $10$ \citep[see][]{Fiore2003}.
An $X/O$ ratio of above 10 is typical of obscured AGNs at high-z and high-luminosity (type 2 QSOs), as
well as high-z clusters of galaxies and extreme BL Lac objects. Values of $X/O$ below $0.1$ are found in coronal emitting stars,
normal galaxies (both early-type and star-forming), and nearby heavily absorbed AGNs \citep{DellaCeca2004}.\\

\begin{figure}[!Hhbt]
        \centering
        \hspace*{-0.2cm}\includegraphics[width=0.50\textwidth]{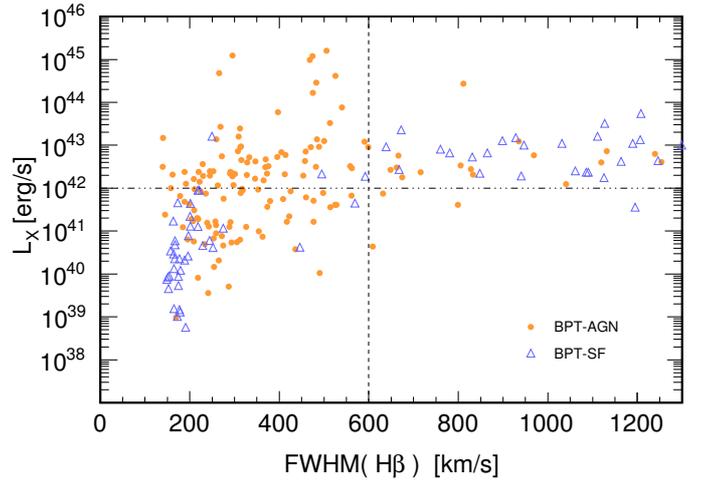}\\
        \caption{X-ray luminosity as a function of the H$_{\beta}$ FWHM. The vertical 
        dotted line marks the threshold of FWHM(H$_{\beta}$)=600 km/s, while the horizontal dashed line corresponds to
	    $L_{X}=10^{42}$ erg/s. Different symbols are related to the optical classification: BPT-AGN as filled circles
        and BPT-SF as empty triangles.}
        \label{fig:LxvsFWHM}
\end{figure}

Thus, we should expect that the thickness parameter and the $X/O$ values fall within the typical range of values for 
BPT-SF galaxies, i.e. $T < 1$ and $X/O < 0.1$ (respectively). We expect correspondingly that BPT-AGN will fall
outside of this range. Figure~\ref{fig:HR2_vs_T_and_XO} shows the combined information provided by these three parameters:
$X/O$ versus $HR$ and $T$ versus $HR$. We have used different symbols to denote different ranges of \lx: 
filled and empty symbols denote the optical classification, identified as BPT-AGN and BPT-SF respectively. 
Whilst nearly all of the X-ray-based AGNs (missing-AGN and strong-AGN subsample, see Table~\ref{tab:subsamples}) 
have typical AGN values for both $X/O$ and $T$ parameters, the values for the true-SF and weak-AGN subsample are more 
consistent with being SF galaxies, for which $X/O$ and $T$ are lower than $0.1$ and $1$,
respectively.  Despite this, we cannot define a range of values that isolate AGN from the rest 
of the sources, either by using $T$ or $X/O$, i.e. a NELG classified as either a SF galaxy or an
AGN (by using either optical-based or X-ray-based criteria) does not occupy a definite region in these parameter spaces.
Analogously, hardness ratios do not clearly cluster around different values for different subclasses of objects. In general,
that one would expect SF galaxies to have an X-ray spectrum dominated by a thermal component that is softer than the
typical power-law spectra exhibited by an AGN. However, the mix of BPT-SF and BPT-AGN galaxies do not show a clear trend
in their hardness ratios. Thus, we cannot establich a clearly defined criterion to differentiate AGN from SF, 
in the context of the three analysed parameters.\\
However, we find that the log $T$ as well as the log $X/O$ distributions 
display bimodal shapes for the BPT-SF population opening the possibility that the emission 
of the missing-AGN population has a different nature (see Figure~\ref{fig:HR2_vs_T_and_XO}).\\
Figure~\ref{fig:LxvsFWHM} shows the X-ray luminosity as a function of the \Hb~ FWHM for these 
two optical populations. From this figure, it is evident that this bimodal feature of the BPT-SF population is directly linked to the 
values of \Hb~ FWHM. There is an almost one-to-one correspondence for the NELGs diagnosed as 
BPT-SF galaxies, between the FWHM of their \Hb~ line and the X-ray luminosity. Among this sample of 66 BPT-SFs, 
we indeed found that roughly all these with \lx$<10^{42}$ \lum~ exhibit an \Hb~ FWHM$\lesssim 600$ km~s$^{-1}$,  
while all the more X-ray luminous objects (which should contain an AGN) have \Hb~ FWHMs of between 600 and 1200 km~s$^{-1}$. 
However, this division based on \Hb~ FWHM does not apply to the BPT-AGNs, where the vast majority are smaller than $\sim$600 
km~s$^{-1}$, independent of the value of \lx.\\

\subsection{An overview of the missing-AGN subsample}

On the basis of the estimated upper limits to the $2-10$ keV luminosity given by FLIX\footnote{FLIX: upper limit server 
for XMM-Newton data provided by the XMM-Newton Survey Science Centre.} for objects that were identified as NELGs
but lacked X-ray detections, we identified another $1207$ galaxies that were classified as SF based on their position 
in the BPT diagram. However, only $5\%$ ($60/1207$) of them could be missing-AGN candidates, i.e. those for which the upper 
limit to their $2-10$ keV exceeds $10^{42}\, {\rm erg}\,{\rm s}^{-1}$.\\
Therefore the missing-AGN candidates represent only between $2\%$ and $7\%$ $(28+60)$, of the BPT-SF population $(66+1207)$
and therefore they do not represent a major problem in terms of \emph{incompleteness}. 
In terms of the total sample of NELGs covered by X-ray observations, the missing-AGN represent between $1.6\%$ and $5\%$ 
of the entire sample. However, the nature of the missing-AGN subsample is poorly understood and needs to be explored further.\\

In the above section we have described the X-ray and optical spectral properties used to explore 
the nature of the elusiveness of optical signatures in the misssing-AGN subsample.  
Similar studies have been performed previously focussing on the nature of the so-called \emph{elusive} AGNs, 
i.e. sources that show no signs of AGN activity in the optical regime, 
but display signs of AGN activity in the X-ray band. One possibility is that they could be mildly/heavily 
obscured AGNs in which star formation dominates the optical emission-line ratios. The obscuration of 
the narrow-line region may well be caused by gas and dust close to the galactic nucleus and 
therefore, comparing the X-ray to [OIII] emission should reveal the level of absorption 
\citep{Maiolino1999,Bassani1999}. However, given that the thickness parameter is in the range $T\geq10$,
the hydrogen column density would have to be $< 10^{23}$ cm$^{-2}$ and therefore obscuration by the torus is not
very likely to be the cause of the elusiveness of optical AGN signatures in these sources.\\

\begin{figure}[!Hhbt]
    \centering
    \includegraphics[width=0.38\textwidth]{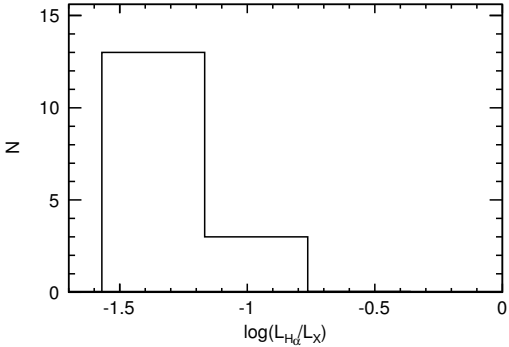}\\
    \caption{Distribution of the $L_{H_{\alpha}}/L_X$ for missing-AGN sub-sample.}
    \label{fig:HbLxratio}
\end{figure}

Another possibility is that the sample contains composite objects, i.e. those hosting both
star formation and an AGN. On the basis of the relation between $L_X(2-10$~keV$)$ and $L_{H_{\alpha}}$ for  
SF galaxies \citep{Ranalli2003,Kennicut1998}, $L_{H_{\alpha}}/L_X$ should be greater than $1$, assuming 
$A_v\leq2$. Whilst type 1 AGNs have ratios lower by two orders
of magnitude, composite galaxies are expected to have intermediate ratios \citep{Yan2011}. Figure~\ref{fig:HbLxratio} shows
the distribution of $\log{(L_{H_{\alpha}}/L_X)}$, where most of the $L_{H_{\alpha}}/L_X$ ratios seem to lie in-between 
both extremes as expected in the composite galaxy range. This opens the possibility that the missing-AGN population 
could be composite objects having both star formation and active nuclei, although largely consistent with being AGN.

\begin{figure}[!Hhb]
    \centering
    \includegraphics[width=8cm]{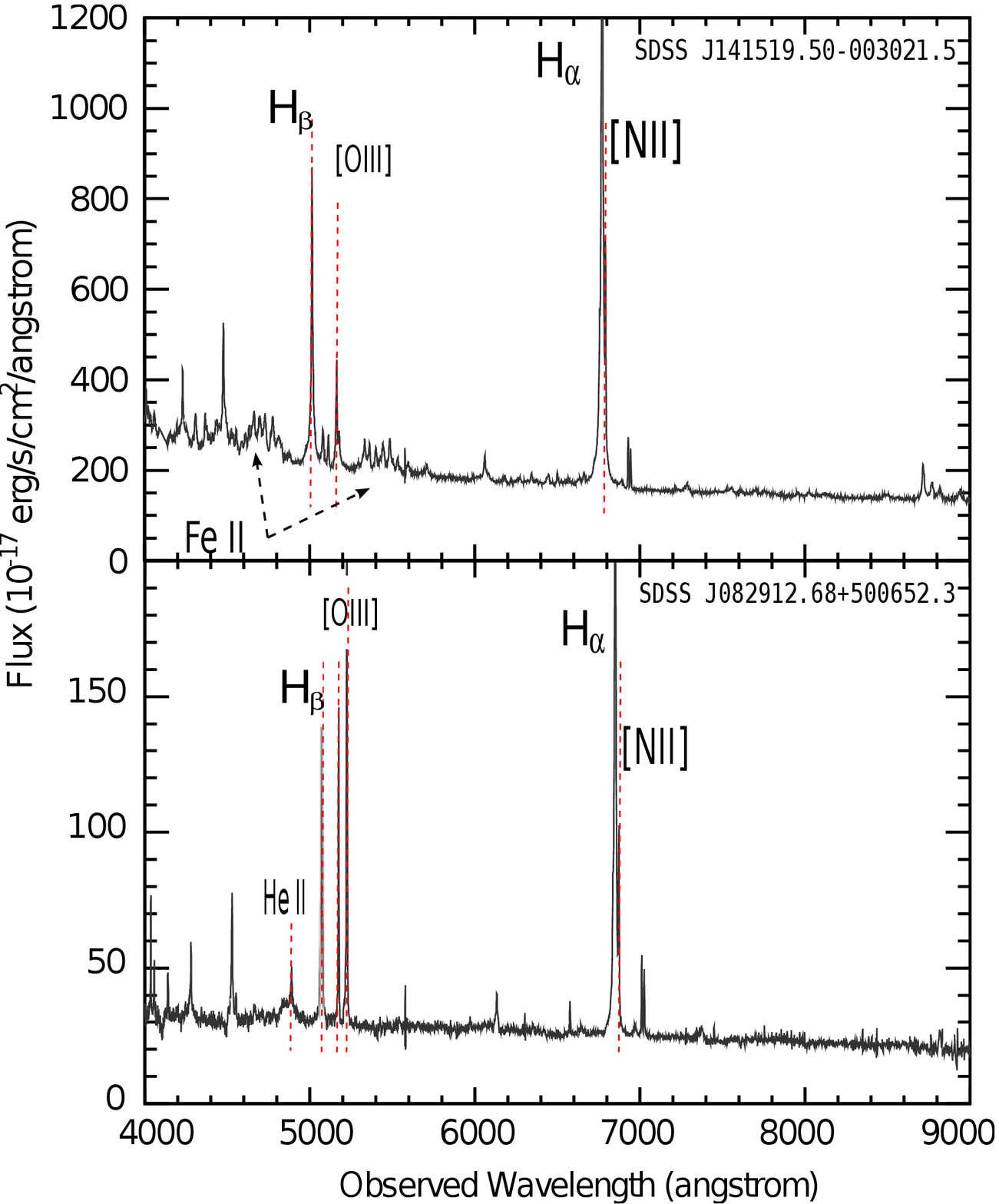}
    \caption{Example of a typical NLS1 with a moderate Fe II emission (\emph{top panel}) and, 
    on the other hand, a SDSS spectrum of an Fe II-lacking NLS1 (\emph{bottom panel}).
    Note in the last spectrum the very narrow Balmer emission lines and also the very weak Fe II multiplets;
    the prominent He II broad emission line ensures our classification as a type 1 AGN.}
    \label{fig:undetectableFeII}
 \end{figure}

\begin{figure}[!Hhbt] 
    \centering
    \includegraphics[width=0.38\textwidth]{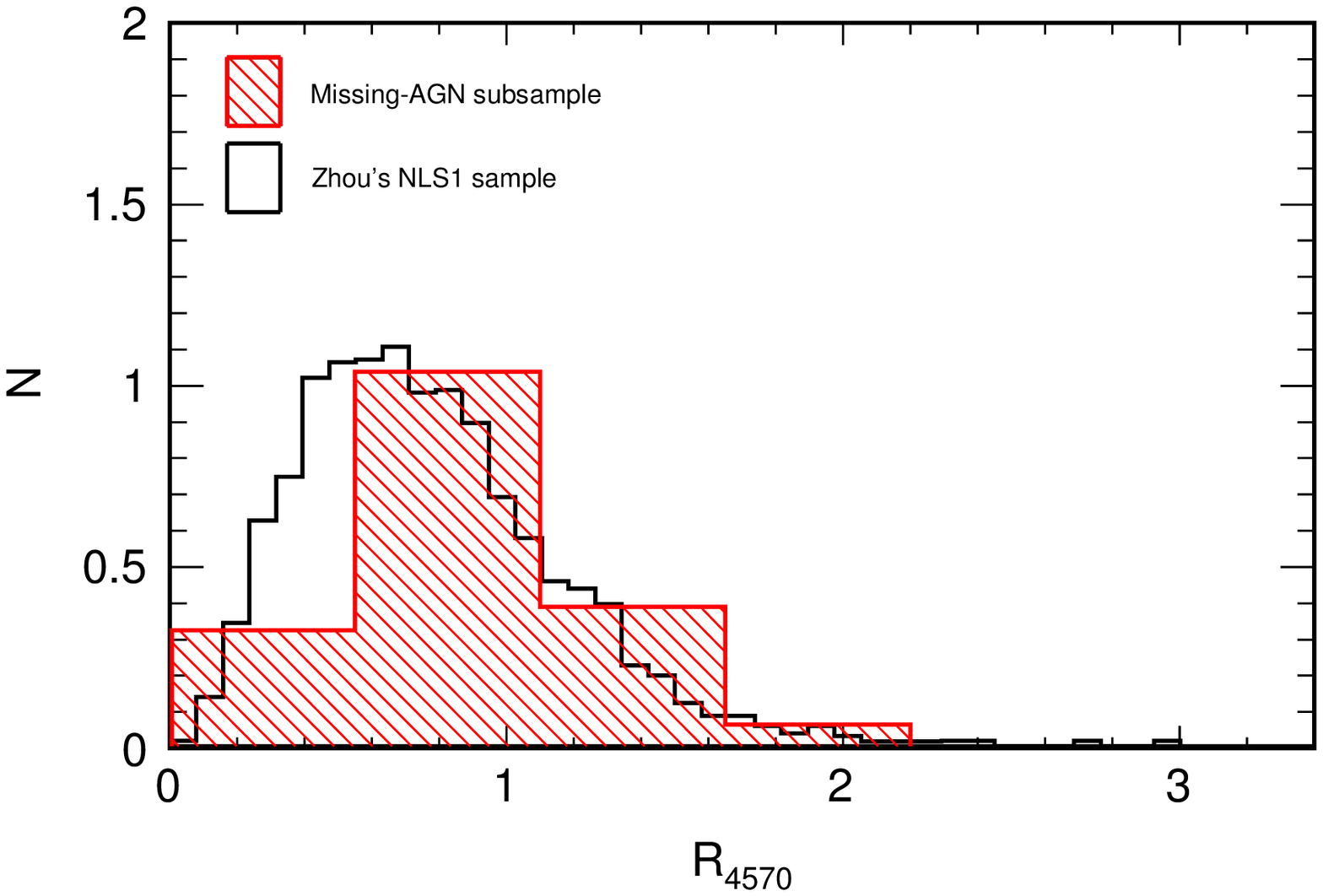}\\
    \caption{Normalized distribution of the relative strength of the Fe II multiplets, $R_{4570}$, for the missing-AGN subsample 
    (filled histogram) and the NLS1 sample of \citet{Zhou2006}.}
    \label{fig:R4570distribution}
\end{figure}

The high values of both $X/O$ and $T$, which are two orders of magnitude higher than the average in SF galaxies
and those typical of Seyfert 1, the low hardness ratio and the quite high values for the \Hb~ FWHM make our missing-AGN sources
very likely to be narrow line Seyfert 1 (NLS1), which are believed to lie in the starburst region of the BPT diagrams.
The NLS1 galaxies represent a subclass \citep{Osterbrock1985} of type 1 AGNs that manifest a distinctive 
ensemble of properties. They are AGN with optical spectral properties similar to those of Seyfert 1 galaxies, except for
recombination lines that are only slightly broader than forbidden emission lines. Studies of NLS1s have idenfified many 
peculiar properties that extend well beyond a pure line-width-based distinction. 
Distinctive features in optical spectra of NLS1s, are the low values of the [OIII]/\Hb, and the often
strong permitted blended FeII emission. Out of the $28$ missing-AGNs, the Fe II multiplets were detected in $23$ objects 
at the $>3\sigma$ level. Among the rest of the subsample, there is another rare class of NLS1s that do not exhibit strong  
Fe II multiplets. For the three objects with very narrow Balmer lines 
(see Figure~\ref{fig:undetectableFeII}), there is a prominent He II broad emission line, which ensures their
classification as type 1 AGNs. For two additional sources, the SDSS spectra were too noisy to yield
reliable measurements of either Fe II multiplets or He II, and in addition they have evidence of high reddening. \\
The relative strength of the Fe II multiplets is usually expressed as the flux ratio of Fe II 
to \Hb: $R_{4570}\equiv FeII\lambda\lambda4434-4684/H_{\beta}$, where Fe II $\lambda\lambda4434-4684$ denotes
the flux of the Fe II multiplets integrated over the wavelength range of $4434-4684$ $\AA{}$ after 
subtracting the local underlying continuum and the He II $\lambda4686$ emission line.
Figure~\ref{fig:R4570distribution} shows the distribution of the relative strength of the Fe II multiplets, $R_{4570}$,
for the missing-AGN subsample, which is compared with the distribution given by \citet{Zhou2006}. 
The average is $<R_{4570}>=0.88$ and the $1$ $\sigma$ dispersion is $0.5$, which is consistent with the Zhou's NLS1 sample:
the probability that both distributions come from the same distribution is $\sim$89\% according to the Kolmogorov-Smirnov test. 
The average is significantly larger than the typical value of $R_{4570}\sim 0.4$ found in normal AGNs \citep{Bergeron1984},
bolstering again the idea that these two populations are probably of a different nature.

\section{X-ray spectral analysis}
\label{section3}
We carried out an X-ray spectral analysis of the BPT-SF populations, consisting of the two subsamples of
missing-AGN and true-SF subsamples. We recall that our main aim is to understand the nature of the
missing-AGN subsample, which are optically classified as SF but have \lx~ in excess of $10^{42}$ \lum~ 
that are indicative of an AGN. The results of the previous section ($T$, $X/O$, $HR$, \Hb~ FWHM, and $R_{4570}$) 
lead us to propose that these objects are good candidate NLS1s. Thus, the missing-AGN and true-SF subsamples 
were analysed as samples of different natures.

\subsection{Data reduction and spectral analysis} 

All objects presented here were observed with \xmm between 2001 June and 2007 December. The European Photon
Imaging Cameras (EPIC) pn \citep{Struder2001} and MOS \citep[MOS1 and MOS2]{Turner2001} were operated in
full frame imaging mode during all the observations. The \xmm data of some objects were previously presented
in the literature (see Table~\ref{tab:MissingAGN}). For a fully homogeneous analysis enabling robust conclusions, we 
reanalysed the \xmm spectra of these objects, in exactly the same way as for the objects whose \xmm data are presented here
for the first time.\\

\begin{table*}[!Hht]
	\caption[]{Summary of some observational parameters.}
	\label{tab:MissingAGN}

\centering
\begin{tabular}{lccccccccc}\hline\hline
ID &  SDSS DR7 & \multicolumn{2}{c}{2XMMi Catalogue}          &  $t_{exp}$ &   EPIC-pn   &  Redshift    &   $N_{H,Gal}$   &   $L_X$ & Notes \\\cline{3-4}
   & SDSS\dots & 2XMM\dots & Obs ID  & ks & Counts$^{3}$ & & $\times 10^{20}$ cm$^{-2}$ & $\times 10^{42}$\lum & \\\hline
&&&&\\
 \multicolumn{10}{c}{``Missing AGN'' sample} \\
&&&&\\\hline
26   &  J141519.49-003021.5 &    J141519.4-003021   &   0145480101  &      13&   1781 $\pm$ 44   &   0.135   &   3.28   &   8.27 & a\\
42   &  J010712.03+140844.9 &    J010712.0+140844   &   0305920101  &      16&   4502 $\pm$ 69   &   0.077   &   3.41   &   2.81 & b\\
53   &  J135724.52+652505.9 &    J135724.5+652506   &   0305920501  &      1.7&   931 $\pm$ 31   &   0.106   &   1.38   &   6.85 & b \\
63   &  J111031.61+022043.2 &    iJ111031.6+022043   &  0504101801  &      8 &   344 $\pm$ 19   &   0.080   &   3.73   &   2.30 &\\
65   &  J114008.71+030711.4 &    J114008.7+030710   &   0305920201  &      34&   23860 $\pm$ 156   &   0.081   &   1.93   &   3.91 &b\\
71   &  J124635.24+022208.7 &    J124635.3+022209   &   0051760101  &      4 &   31746 $\pm$ 179   &   0.048   &   1.85   &   12.51 &c\\
100  &  J221918.53+120753.1 &   J221918.5+120753   &   0103861201  &      8  &   19684 $\pm$ 141   &   0.081   &   5.03   &   10.03 &d\\
104  &  J092247.02+512038.0 &   J092246.9+512037   &   0300910301  &      6.3&   12029 $\pm$ 111   &   0.160   &   1.32   &   18.90 &\\
111  &  J094240.92+480017.3 &   J094240.9+480017   &   0201470101  &      14 &   424 $\pm$ 24   &   0.197   &   1.20   &   5.87 &\\
125  &  J133141.03-015212.4 &   J133141.0-015212   &   0112240301  &      24 &   2548 $\pm$ 52   &   0.145   &   2.39   &   11.90 &\\
129  &  J081053.75+280610.9 &   J081053.8+280611   &   0152530101  &      16 &   986 $\pm$ 33   &   0.285   &   2.93   &   18.17 &\\
161  &  J123126.44+105111.3 &   J123126.4+105111   &   0145800101  &      7.7&   269 $\pm$ 18   &   0.304   &   2.14   &   6.86 &\\
$^{\star}$163  &  J123748.49+092323.1 &   iJ123748.5+092323  &   0504100601  &             &                  & 0.125 & 1.48 & 		    & \\
171  &  J093922.90+370943.9 &   J093922.9+370942   &   0411980301  &      4   &   2857 $\pm$ 54   &   0.186   &   1.22   &   27.89 &\\
191  &  J155909.62+350147.4 &   J155909.6+350147   &   0112600801  &      11  &   80531 $\pm$ 285   &   0.031   &   2.11   &   8.22 &\\
195  &  J103438.59+393828.2 &   J103438.6+393828   &   0109070101  &      28  &  21857$\pm$214   & 0.043  & 1.31  &        & c,e\\
203  &  J124013.82+473354.7 &   J124013.8+473355   &   0148740501  &      5.7 &   804 $\pm$ 29   &   0.117   &   1.32   &   1.03 &\\
204  &  J124058.45+473302.0 &   J124058.3+473302   &   0148740501  &      5   &   192 $\pm$ 15   &   0.367   &   1.33   &   35.88 &\\
214  &  J112405.15+061248.8 &   J112405.1+061248   &   0103863201  &      5   &   786 $\pm$ 29   &   0.272   &   4.61   &   36.76 &\\
241  &  J075216.55+500251.3 &   J075216.4+500251   &   0151270201  &      7.7 &   620 $\pm$ 26   &   0.263   &   5.17   &   57.45 &\\
275  &  J145108.76+270926.9 &   J145108.7+270926   &   0152660101  &      18  &   91984 $\pm$ 305   &   0.065   &   2.70   &   20.35 &f\\
302  &  J102812.67+293222.8 &   J102812.6+293222   &   0301650401  &      8.4  &   76 $\pm$ 13   &   0.287   &   1.91   &   3.57 & \\
318  &  J122230.71+155547.9 &   J122230.7+155547   &   0106860201  &      8.6  &   238 $\pm$ 17   &   0.367   &   1.99   &   16.00 &\\
329  &  J140621.89+222346.5 &   J140621.8+222347   &   0051760201  &      3.1  &   1911 $\pm$ 44   &   0.098   &   2.05   &   3.24 & c,g\\
$^{\star}$335  &  J014856.95+135451.8 &   J014856.9+135450   &   0094383401  &             &                   &   0.220   &  4.90    &        & \\
338  &  J082912.67+500652.3 &   iJ082912.8+500652  &   0303550901  &      2.2   &   442 $\pm$ 22   &   0.043   &   4.07   &   2.51 &b\\
355  &  J131718.58+324035.6 &   J131718.6+324036   &   0135940201  &      10   &   161 $\pm$ 13   &   0.061   &   1.17   &   2.27 &\\
357  &  J134235.66+261534.0 &   J134235.6+261534   &   0108460101  &      26   &   867 $\pm$ 30   &   0.064   &   1.03   &   2.59 &\\\hline
&&&&&\\
 \multicolumn{10}{c}{Sub-sample of optically-classified SF galaxies} \\
&&&&\\\hline
8    & J140919.94+262220.1 &  J140920.0+262219 & 0092850501 & 35 & $163\pm20$ &  0.059 &  1.40 & 6.0   \\
56   & J095848.66+025243.2 &  J095848.6+025243 & 0203362101 & 54 & $259\pm32$ &  0.079 &  1.83 & 36.0  \\
79   & J093402.02+551427.8 &  J093401.9+551428 & 0112520101 & 27 & $2178\pm56$&  0.002 &  2.46 & 0.2   \\
154  & J162636.40+350242.0 & iJ162636.5+350242 & 0505011201 & 14 & $127\pm12$ &  0.034 &  1.44 & 4.2   \\
164  & J080629.80+241955.6 &  J080629.7+241956 & 0203280201 & 6  & $156\pm21$ &  0.042 &  3.80 & 45.0  \\
233  & J122254.57+154916.4 &  J122254.6+154916 & 0106860201 & 10 & $473\pm32$ &  0.005 &  2.01 & 0.8   \\
246  & J085735.33+274605.1 &  J085735.4+274607 & 0210280101 & 68 & $256\pm18$ &  0.007 &  2.51 & 0.2   \\
251  & J123520.04+393109.1 &  J123519.9+393110 & 0204400101 & 26 & $97\pm8$   &  0.021 &  1.31 & 0.5   \\ \hline \hline
\end{tabular}
\tablefoot{\emph{Left to right:} Numeric identifier of the source, SDSS object name where the full name should be `SDSS \dots', 
2MM object name where the full name should be
`2XMM J\dots', XMM-Newton's observation number,  XMM-Newton's exposure time in units of ks, the total counts in the EPIC pn monitor, 
Galactic column density from \citet{Dickey1990} in units of $10^{20}$ cm$^{-2}$, and the hard X-ray luminosity. 
The last column give references about its classification for some sources: 
(a) \citet{26_Foschini2004}, (b) \citet{42_Dewangan2008}, (c) \citet{71_Piconcelli2005},
(d) \citet{100_Gallo2006}, (e) \citet{195_Maitra2010}, (f) \citet{275_Grupe2010},
(g) \citet{329_Crummy2006} (among others). \\
$^{\star}$ These sources could not be analysed owing to the unreliability quality of the data statistics 
(number of EPIC-pn counts less than $50$).
}
\end{table*}

We chose to use EPIC-pn data as it covers a larger effective area resulting in a higher S/N. The observation data
files (ODFs) were processed to produce calibrated event lists using the Science Analysis System (SAS 10.0.0). 
We extracted the source spectra using the good EPIC-pn events in circular regions of radii ranging from 
12$^{''}$ to 30$^{''}$ centred on the source position.
We used single- and double-pixel events for all observations. The background spectra were extracted from nearby circular regions
free of sources. Spectral response files were generated using the SAS tasks {\tt rmfgen} and {\tt arfgen}. The {\tt epatplot} SAS task
was used to test for the presence of pile up. The EPIC-pn  X-ray spectra of all but two of the observations (sources 2XMM
J124635.3+022209 and 2XMM J103438.6+393828, see Table~\ref{tab:MissingAGN}) were found to be free 
from the effects of pile-up. We performed an X-ray
spectral analysis with XSPEC v12.5 \citep{Arnaud1996}, taking the limits in accurate calibration of the pn data
as $0.3-12$ keV. The source spectra were grouped to have at least 20 counts in each bin in order to apply the modified $\chi^2$
minimization technique; the lowest quality spectra observations ($100$-$300$ counts) were only grouped with
at least $15$ counts per bin. We did not carry out an X-ray spectral analysis for the faintest sources ($<100$). All quoted
errors are for a 90 per cent confidence interval for one parameter ($\Delta\chi^2=2.706$).\\

The X-ray selection criteria
resulted in a minimum of 30 counts at energies above 2 keV in at least one detector independently of the quality of the spectrum, 
which means that the S/N was sometimes low. Thus, the level of detail of our spectral analysis varied for each source depending on the
quality of the EPIC-pn spectra, ranging from a quite detailed analysis for bright sources, to only very coarse spectral fits for
the faintest. 

\subsection{Missing-AGN subsample} 
We now study the nature of the missing-AGN subsample, which are possibly NLS1s. 
Two objects (2XMMi J123748.5+092323 and 2XMM J014856.9+135450, see Table~\ref{tab:MissingAGN}) 
simply could not be modelled owing to the low number of counts ($<$50). 
For three other sources (2XMM J141519.4-003021, 2XMM J135724.5+652506, and 2XMM J123126.4+105111, 
see Table~\ref{tab:MissingAGN}) the energy range used by the spectral analysis was E$\lesssim4$ keV
due to the EPIC-pn data being dominated by the background above these energies. Finally, 2XMMi J082912.8+500652,
2XMM J131718.6+324036, and 2XMM J134235.6+261534 were analysed using only EPIC-MOS data because of the lack of EPIC-pn data.
In Table~\ref{tab:MissingAGN}, we give details of the X-ray observations and the hard X-ray luminosity of each object, 
which was calculated using the best-fit power-law model over the \he~ keV energy band. We note that the Galactic absorption is implicitly 
included in all the spectral models presented hereafter at the values given in Table~\ref{tab:MissingAGN_fit_pl}.\\

The general best-fit model of the X-ray spectrum emitted by a NLS1 has typically four components: an underlying absorbed
steep power-law, a soft X-ray excess, and a reflection component that might also include a broad feature near the
Fe line complex at $5-7$ keV. Several explanations have been proposed for the origin of the observed soft excess, such as a
relativistically blurred photoionized disc reflection \citep{Ross2002,Crummy2006},
an intrinsic thermal component, or ionized absorption arising in a wind from the inner disc 
\citep{Gierlinski2004,Gierlinski2006}. 
For simplicity, we only used two different two-component continua: a partial covering 
and a thermal model as proxies to each explanation respectively (the
quality of the X-ray data did not allow a more sophisticated analysis in the majority of cases). 
In the case of the first two-component continua, we modelled the soft excess emission due to reprocessing of the 
primary X-rays as a partial-covering neutral material (PCF model). This can be regarded as a physical model 
of a clumpy torus, where the torus is a smooth continuation of the broad line region, and provide a physical 
explanation of the apparent mismatch between the optical classification and the X-ray properties of these objects. 
In this model, X-ray absorption, dust obscuration, and broad line emission are produced in a single continuous 
distribution of clouds: the broad line region is located within the dust sublimation radius, hence the non-dust-free 
clouds obscure the optical emission but not the X-rays, whereas the torus is located outside the dust sublimation ratio. 
The PCF model assumes that some fraction, $f$, of the X-ray source is covered by a neutral absorber with a column 
density of $N_{H,z}$, while the rest is unobscured. This could be responsible for an apparent soft excess in two different 
geometries, either by reflection from optically thick material out of the line of sight \citep{Fabian2002}, or absorption by
optically thin material along the line of sight \citep{Gierlinski2004,Chevallier2006}.\\

On the other hand, there are some possible ways of explaining the soft excess from the disc itself in terms of the 
reprocessing of the
primary X-rays in the accretion disc as reflected emission from a geometrically flat disc, with solar abundances, 
illuminated by an isotropic source. Thus, the soft excess is sometimes closely fitted by a black body that has 
a roughly constant temperature of $0.1-0.2$ keV. If this radiation is thermal, this temperature is much too 
high to be explained by the standard accretion disc model of \citet{Shakura1973}, although it could be explained
by a slim accretion disc in which the temperature is raised by photon trapping, in which case the accretion is 
super-Eddington \citep{Tanaka2005}, or by thhe Comptonization of extreme UV accretion disc photons \citep[][e.g.]{Porquet2004}.
\\

We carried out the {\bf hard X-ray spectral fitting procedure} using the following scheme:
\begin{enumerate}
 \item We first fitted the individual data in the \he~ keV (when possible) energy range, using a power law modified
 by absorption from  cold gas in our Galaxy (hereafter PL).
 \item We then checked for any significant additional component  that may be present in this energy range, such as an iron
 emission line (model led with {\tt zgauss} at  $E_c\sim6.4$ keV) and/or an Fe K-edge among other features.
\end{enumerate}

\begin{table*}[!htb]
	\caption{\emph{Missing AGNs}. Results of fitting the spectral data with both a power law in the hard X-ray energy band
	($2-10$ keV) and an absorbed power-law model across the whole X-ray spectrum ($0.3-10$ keV).}
	\label{tab:MissingAGN_fit_pl}
\centering
\begin{tabular}{lccc}\hline\hline
ID  & Notes & $\Gamma$ 	& $\chi^2_{\nu}/$d.o.f.  \\\hline
&&& \\
\multicolumn{4}{c}{Pawer-Law model over $2-10$ keV} \\
&&& \\\hline
26 	& 	& $1.50\pm^{0.72}_{0.73}$ 	& $1.020/3$\\      
42	& 	& $2.23\pm^{0.25}_{0.32}$ 	& $0.680/20$\\
53	& $a$ 	& $2.75\pm^{0.43}_{0.37}$ 	& $1.306/11$\\
63	& $a$	& $2.20\pm^{0.53}_{0.45}$ 	& $0.829/7$\\
65	& 	& $2.27\pm^{0.14}_{0.19}$ 	& $1.156/54$\\
71      & $c$   & $2.90\pm^{0.02}_{0.02}$ 	& $2.080/112$\\
100	& 	& $2.38\pm^{0.16}_{0.15}$ 	& $1.182/47$\\
104	& 	& $2.24\pm^{0.40}_{0.51}$ 	& $0.581/9$\\
111	& $b,1$ &  				& \\
125     & $b,2$ & $1.71\pm^{0.41}_{0.61}$ 	& $0.655/6$\\
129 	& 	& $2.31\pm^{0.77}_{0.74}$ 	& $1.270/3$\\
161     & 	& 				&\\
171 	& 	& $2.19\pm^{0.56}_{0.63}$ 	& $0.811/3$\\
191     &       & $2.11\pm^{0.06}_{0.06}$ 	& $1.195/217$\\
195     &	& $2.10\pm^{0.02}_{0.03}$	& $1.021/299$\\
203 	& 	& $1.68\pm^{0.50}_{0.66}$ 	& $0.817/4$\\
204 	& $a$	& $2.03\pm^{0.73}_{0.66}$ 	& $0.220/3$\\
214 	& 	& $1.64\pm^{0.51}_{0.50}$ 	& $0.958/4$\\
241 	& 	& $1.28\pm^{0.63}_{0.59}$ 	& $0.810/19$\\
275	& 	& $2.75\pm^{0.07}_{0.07}$ 	& $0.897/213$\\
318	& $b,1$ &  	&\\
329 	& $c$   & $1.74\pm^{0.97}_{0.90}$ 	& $1.910/3$\\
338     &  	& $2.20\pm^{0.60}_{0.93}$ 	& $0.802/7$\\
355     &  	& $1.57\pm^{0.30}_{0.52}$ 	& $0.008/1$\\
357     &  	& $1.83\pm^{0.38}_{0.40}$ 	& $0.556/11$\\\hline\hline
\end{tabular}
\hspace*{1cm}
\begin{tabular}{lccc}\hline\hline
ID  & Notes & $\Gamma$ 	& $\chi^2_{\nu}/$d.o.f.  \\\hline

&&& \\
\multicolumn{4}{c}{Absorbed Power-Law model over $0.3-10$ keV} \\
&&& \\\hline
26 	&	& $2.79\pm^{0.09}_{0.09}$ 	& $1.213/63$\\
42	& 	& $2.39\pm^{0.05}_{0.05}$ 	& $0.933/161$\\
53	& $a$ 	& $2.57\pm^{0.11}_{0.10}$ 	& $1.154/43$\\
63	& $a$	& $1.91\pm^{0.19}_{0.17}$ 	& $1.500/18$\\
65	& 	& $2.79\pm^{0.02}_{0.02}$ 	& $1.370/316$\\
71      & $c$   &  &  \\
100	& 	& $2.95\pm^{0.03}_{0.02}$ 	& $1.360/301$\\
104	& 	& $3.48\pm^{0.04}_{0.04}$ 	& $1.350/170$\\
111	& $b,1$ & $2.72\pm^{0.25}_{0.24}$ 	& $0.676/31$\\
125     & $b,2$ & $2.58\pm^{0.08}_{0.07}$ 	& $0.993/85$\\
129 	& 	& $2.31\pm^{0.12}_{0.11}$ 	& $1.294/39$\\
161     & 	& $3.17\pm^{0.34}_{0.32}$ 	& $1.762/8$\\
171 	& 	& $2.95\pm^{0.08}_{0.09}$ 	& $1.437/81$\\
191     &       & $2.67\pm^{0.01}_{0.01}$ 	& $1.475/562$\\
195     &	& $2.10\pm^{0.02}_{0.03}$	& $2.021/587$\\
203 	& 	& $2.24\pm^{0.12}_{0.12}$ 	& $0.949/41$\\
204 	& $a$	& $2.45\pm^{0.35}_{0.31}$ 	& $0.367/13$\\
214 	& 	& $2.70\pm^{0.14}_{0.14}$ 	& $1.052/32$\\
241 	& 	& $2.37\pm^{0.18}_{0.17}$ 	& $0.899/102$\\
275	& 	& $2.85\pm^{0.02}_{0.02}$ 	& $1.657/558$\\
318	& $b,1$ & $2.77\pm^{0.39}_{0.35}$ 	& $0.894/30$\\
329 	& $c$   &   &\\
338     &  	& $2.38\pm^{0.09}_{0.09}$ 	& $0.979/63$\\
355     &  	& $1.65\pm^{0.20}_{0.19}$ 	& $1.002/13$\\
357     &  	& $2.05\pm^{0.08}_{0.08}$ 	& $1.231/66$\\\hline\hline
\end{tabular}
\centering
\tablefoot{
(a) The hard photon index was obtained in the energy band $1-10$ keV due to low statistics;\\
(b) The \emph{pn} data is dominated by the background: (1) above $2-3$ keV, (2) above $6$ keV;\\
(c) The soft excess is far strong (\rm{ratio} $> 30$) invalidating the application of PL fit across the whole X-ray spectra.\\
}
\end{table*}

\begin{figure}[!Hht] 
\centering
        \includegraphics[width=0.5\textwidth]{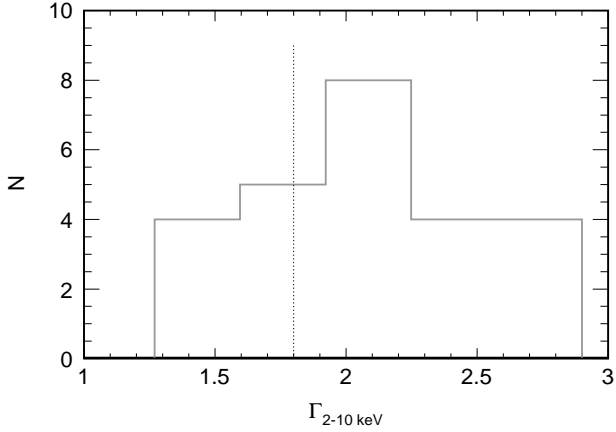}
        \caption{Distribution in $\Gamma_{2-10}$ for the ``missing AGN'' when we fitted the individual data in the
        $2-10$ keV energy range with a single power law modified by absorption in our Galaxy. The vertical line
        marks the expected value for objects of Seyfert 2 type.}
        \label{fig:Gamma}
\end{figure}

We found that the PL yields an acceptable fit for almost all objects in the sample, in terms of the minimum $\chi^2$ 
(see Table~\ref{tab:MissingAGN_fit_pl}). No other statistically significant and physically meaningful 
features were found in the hard X-ray spectra. Figure~\ref{fig:Gamma} shows that the 
distribution of $\Gamma_{2-10}$ takes values  above the typical expected value ($\sim 1.8$) 
found for type 2 AGNs. We note that the cases where the
X-ray spectra appears very hard with a power-law slope $\Gamma\sim1.5$, also correspond to those 
with larger errors, $\pm0.5$ (see Table~~\ref{tab:MissingAGN_fit_pl}).\\
We also attempted to fit the spectrum over the entire X-ray band ($0.3-10$ keV) with an 
intrinsically absorbed power-law model (absorbed-PL). We found that for all sources this simple model 
was rejected over the whole X-ray band with a high statistical significance. 
The estimated intrinsic column density was found to be less than $10^{22}$ cm$^{-2}$, which corresponds to an 
unabsorbed model or at least values of absorption that fall in the low part of the column 
density distribution of type-2 AGNs. These very low upper limits are not quoted in Table~~\ref{tab:MissingAGN_fit_pl}. \\

\begin{figure*}[!Hht]
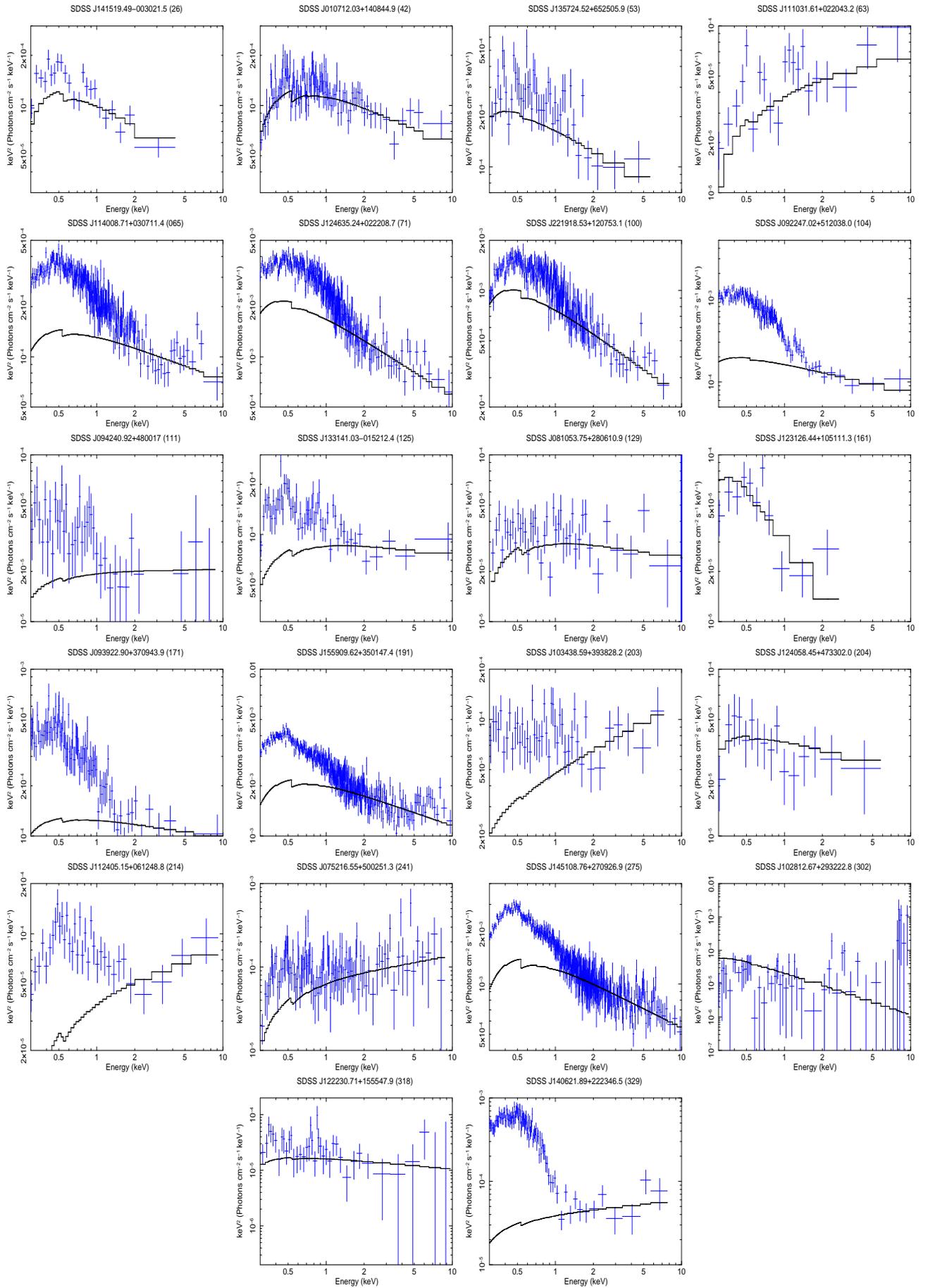

	\centering
	    \includegraphics[width=4.0cm,height=4.2cm,angle=-90]{026_Xray.ps}
	    \includegraphics[width=4.0cm,height=4.2cm,angle=-90]{042_Xray.ps}
        \includegraphics[width=4.0cm,height=4.2cm,angle=-90]{053_Xray.ps}
        \includegraphics[width=4.0cm,height=4.2cm,angle=-90]{063_Xray.ps}
        \includegraphics[width=4.0cm,height=4.2cm,angle=-90]{065_Xray.ps}
        \includegraphics[width=4.0cm,height=4.2cm,angle=-90]{071_Xray.ps}
        \includegraphics[width=4.0cm,height=4.2cm,angle=-90]{100_Xray.ps}
        \includegraphics[width=4.0cm,height=4.2cm,angle=-90]{104_Xray.ps}
        \includegraphics[width=4.0cm,height=4.2cm,angle=-90]{111_Xray.ps}
        \includegraphics[width=4.0cm,height=4.2cm,angle=-90]{125_Xray.ps}
	    \includegraphics[width=4.0cm,height=4.2cm,angle=-90]{129_Xray.ps}
        \includegraphics[width=4.0cm,height=4.2cm,angle=-90]{161_Xray.ps}
        \includegraphics[width=4.0cm,height=4.2cm,angle=-90]{171_Xray.ps}
        \includegraphics[width=4.0cm,height=4.2cm,angle=-90]{191_Xray.ps}
        \includegraphics[width=4.0cm,height=4.2cm,angle=-90]{203_Xray.ps}
        \includegraphics[width=4.0cm,height=4.2cm,angle=-90]{204_Xray.ps}
        \includegraphics[width=4.0cm,height=4.2cm,angle=-90]{214_Xray.ps}
        \includegraphics[width=4.0cm,height=4.2cm,angle=-90]{241_Xray.ps}
	    \includegraphics[width=4.0cm,height=4.2cm,angle=-90]{275_Xray.ps}
        \includegraphics[width=4.0cm,height=4.2cm,angle=-90]{302_Xray.ps}
        \includegraphics[width=4.0cm,height=4.2cm,angle=-90]{318_Xray.ps}
        \includegraphics[width=4.0cm,height=4.2cm,angle=-90]{329_Xray.ps}	
	\caption{The X-ray spaltral fitting and optical data for each object within Missing AGN sample. 
    Objects order on the figure follows the Table 2. X-ray band fitting plot (\emph{Odd panels}): 
    the individual hard ($2-10$ keV) X-ray data has been fitted with a power-law model modified by 
    absorption and we have added the soft component defined as the excess over an extrapolation down 
    up to $0.3$ keV of the best power-law model fitting to the hard X-ray band only for those objects
    where It was made possible by the statistics. \emph{Even panels}: SDSS spectrum plotted as ($\log{F_{\lambda}}$)
    in units of $10^{17}$ erg/cm$^{2}$/s/$\AA{}$. Note that, for two sources ($163$ and $335$) we 
    have not X-ray spectrum due to the low counts in the X-ray energy band.}
	\label{fig:Spectra}
\end{figure*}

We found additional evidence against starburst activity in the missing-AGN population by comparing the soft ($0.3-2.5$ keV) 
versus hard ($2-10$ keV) spectral indices, which revealed an overall spectral steepening towards low energies in many cases. 
This suggests that there is a soft X-ray excess that contributes mostly below $\sim 2$ keV .
For the majority of the sources, this soft excess represents
more than $20\%$ of the X-ray emission, what is higher than expected for starburst activity. 
We show this soft component in Figure~\ref{fig:Spectra}, defined as the excess over an extrapolation to $0.3$ keV 
of the PL model fitted to the hard X-ray band. \\

We carried out the {\bf soft X-ray component fitting procedure} using the following approach and used 
the probability of the F-test, accepting additional  spectral components only when they improved the fit with a significance 
$\geq 3 \sigma$:
\begin{enumerate}
 \item We added either a redshifted black body  component ({\tt zbbody} in XSPEC) 
     to the PL model (hereafter BB-PL model)  or a neutral  absorber 
     at the redshift of the X-ray source that is either fully ($f=1$) or  partially ($f<1$) covers the 
     source ({\tt zpcfabs} in XSPEC; hereafter PCF-PL model).
 \item We then compared these two models\footnote{We note that for most of  the sources the X-ray data quality is 
     too poor to attempt more sophisticated fits.},
     BB-PL and  PCF-PLi. When one of the two models gave a fit with a $\Delta\chi^2\equiv \chi^2_{PL} - \chi^2_{BB-PL/PCF-PL} 
     \geq 10$ and/or an F-test significance that was really high, $\ge 99\%$, the new model was taken as the baseline. 
     In the case of sources for which the $\chi^2$ for the BB-PL and PCF-PL
     models were comparable and the values of each free parameter were physically plausible for both components,
     we adopted as the best fit model the one with the least uncertain model parameters. 
\end{enumerate}

\begin{table*}[!hbt]
\caption{\emph{Missing AGNs}. Best-fit models for the soft X-ray excess.}
	\label{tab:MissingAGN_fit_SE}

\centering

\begin{tabular}{l|cccrc|ccclrc}\hline\hline
ID & $\Gamma$ & KT 		& $\chi^2_{\nu}/d.o.f.$ & $\Delta\chi^2$ & P$_{F-test}^{\dagger}$
   & $\Gamma$   & $N_{H,z}$	& $f$  			& $\chi^2_{\nu}/$d.o.f. & $\Delta\chi^2$  & P$_{F-test}^{\dagger}$ \\ \cline{2-3}\cline{7-9}
   & \multicolumn{2}{c}{ \tiny{\rm{ BB } $\equiv$ \rm{ PHA * ( zPOW + zBB ) }}} & & &
   & \multicolumn{3}{c}{ \tiny{\rm{ PCF } $\equiv$ \rm{ PHA * zPOW * zPCF }}} & & &\\ \hline

\multicolumn{12}{l}{ }\\
\multicolumn{12}{l}{ BB-PL as the best-fit model} \\\hline
65    	& $2.52\pm^{0.04}_{0.03}$ 	& $137\pm^{6}_{6}$ 	& $0.987/314$ 	& $91.4$ & $\sim 0$ 
	& $2.81\pm^{0.03}_{0.03}$	& $49\pm^{35}_{19}$	& $0.7\pm^{0.2}_{0.1}$ 	& $1.171/314$ 	& $33.6$ & $\sim0$ \\
71$^{a}$& $2.50\pm^{0.03}_{0.03}$ 	& $160\pm^{10}_{10}$ 	& $1.140/110$ 	& $108$ & (*)
	& 	& 	&  	&  	&  \\
100     & $2.65\pm^{0.07}_{0.08}$ 	& $160\pm^{10}_{10}$ 	& $1.100/299$ 	& $81.5$ & $\sim0$
	& $2.98\pm^{0.03}_{0.03}$	& $34\pm^{19}_{12}$	& $0.6\pm^{0.1}_{0.1}$ 	& $1.230/299$ 	& $41.5$ & $\sim0$ \\
104     & $2.82\pm^{0.09}_{0.08}$ 	& $119\pm^{4}_{4}$ 	& $1.026/168$ 	& $100.4$ & $\sim0$
	& $3.50\pm^{0.04}_{0.04}$	& $21\pm^{10}_{40}$	& $0.8\pm^{0.1}_{0.2}$ 	& $1.537/168$ 	& $18.9$ & $\sim0$ \\
195$^{a}$& $2.30\pm^{0.06}_{0.13}$ 	& $143\pm^{30}_{21}$ 	& $1.140/210$ 	& $68$ & $\sim0$
	& 	& 	&  	&  	&  \\
241     & $1.79\pm^{0.23}_{0.27}$ 	& $112\pm^{27}_{23}$ 	& $0.764/100$ 	& $15.3$ & $\sim0$
	& $2.62\pm^{0.38}_{0.23}$	& $13\pm^{63}_{10}$	& $0.7\pm^{0.2}_{0.2}$ 	& $0.810/100$ 	& $10.7$ & $\sim0$\\
275     & $2.51\pm^{0.02}_{0.03}$ 	& $116\pm^{2}_{2}$ 	& $1.082/556$ 	& $323.0$ & $\sim0$
	& $2.91\pm^{0.02}_{0.02}$	& $15\pm^{4}_{3}$	& $0.57\pm^{0.03}_{0.04}$ 	& $1.171/556$ 	& $273.5$ & $\sim0$\\
329$^{a}$& $1.65\pm^{0.33}_{0.32}$ 	& $104\pm^{3}_{3}$ 	& $0.844/95$ 	& $214.9$ & (*)
	& 	& 	&  	&  	&  \\ \hline
\multicolumn{12}{l}{ }\\
\multicolumn{12}{l}{ PCF-PL as the best-fit model} \\\hline
26  	& $2.47\pm^{0.11}_{0.27}$ 	& $110\pm^{24}_{42}$ 	& $1.169/61$ 	& $5.1$ & $0.12$
	& $2.84\pm^{0.10}_{0.09}$	& $34\pm^{71}_{19}$	& $0.8\pm^{0.2}_{0.2}$ 	& $1.072/61$ 	& $11.1$  & $\sim0$ \\
42   	& $2.22\pm^{0.05}_{0.04}$ 	& $155\pm^{22}_{21}$ 	& $0.889/159$ 	& $8.9$ & $\sim0$
	& $2.41\pm^{**}_{**}$		& $65\pm^{10}_{51}$	& $0.6\pm^{0.4}_{0.5}$ 	& $0.923/159$ 	& $29.8$ & $0.157$ \\
125     & $2.34\pm^{0.17}_{0.10}$ 	& $102\pm^{25}_{35}$ 	& $0.936/84$ 	& $6.5$ & $\sim0$
	& $2.65\pm^{0.09}_{0.08}$	& $17\pm^{35}_{12}$	& $0.6\pm^{0.3}_{0.2}$ 	& $0.880/84$ 	& $10.5$ & $\sim0$ \\
171     & $3.04\pm^{0.10}_{0.05}$ 	& $>2\cdot10^3$ 	& $0.901/79$ 	& $12.8$ & $\sim0$
	& $3.04\pm^{0.12}_{0.08}$	& $11\pm^{5}_{15}$	& $0.6\pm^{0.2}_{0.2}$ 	& $0.874/79$ 	& $14.9$ & $\sim0$\\
191     & $2.44\pm^{0.03}_{0.03}$ 	& $106\pm^{4}_{4}$ 	& $1.178/560$ 	& $168.4$ & $\sim0$
	& $2.73\pm^{0.01}_{0.01}$	& $23\pm^{7}_{7}$	& $0.6\pm^{0.1}_{0.1}$ 	& $1.041/560$ 	& $245$ & $\sim0$\\
214     & $1.98\pm^{0.20}_{0.15}$ 	& $144\pm^{23}_{27}$ 	& $0.810/30$ 	& $9.3$ & $\sim0$
	& $2.81\pm^{0.15}_{0.15}$	& $43\pm^{86}_{27}$	& $0.9\pm^{0.1}_{0.2}$ 	& $0.681/30$ 	& $13.2$ & $\sim0$ \\\hline

\multicolumn{12}{l}{ }\\
\multicolumn{12}{l}{ BB and/or PCF extra component could not be excleded} \\\hline
53   	& $2.41\pm^{0.26}_{0.20}$ 	& $179\pm^{50}_{41}$ 	& $1.070/41$ 	& $5.7$ & $0.08$ 
	& $2.88\pm^{0.41}_{0.24}$	& $0.05\pm^{0.05}_{**}$	& $0.9\pm^{**}_{**}$ 	& $1.130/41$ 	& $<4$ & $0.25$\\
63      & $2.03\pm^{0.24}_{0.22}$ 	& $>2\cdot10^3$ 	& $1.553/16$ 	& $<3$ & $0.51$
	& $2.01\pm^{0.20}_{0.62}$	& $32\pm^{**}_{**}$	& $0.7\pm^{**}_{**}$ 	& $1.522/16$ 	& $<3$ & $0.44$ \\
129     & $2.12\pm^{0.08}_{0.16}$ 	& $87\pm^{26}_{48}$ 	& $1.195/37$ 	& $6.3$ & $0.08$
	& $2.67\pm^{0.24}_{0.46}$	& $2\pm^{157}_{2}$	& $0.5\pm^{0.2}_{0.3}$ 	& $1.256/37$ 	& $<5$ & $0.22$\\
203     & $2.16\pm^{0.15}_{0.13}$ 	& $<1$		& $0.892/39$ 	& $<5$ & $0.12$ 
	& $2.34\pm^{0.23}_{0.18}$	& $30\pm^{**}_{**}$	& $0.7\pm^{**}_{**}$ 	& $0.935/39$ 	& $<5$ & $0.09$\\
204     & $2.03\pm^{0.63}_{0.48}$ 	& $120\pm^{**}_{**}$ 	& $0.277/11$ 	& $<5$ & $0.08$ 
	& $2.86\pm^{0.70}_{0.60}$	& $4\pm^{**}_{**}$	& $0.7\pm^{**}_{**}$ 	& $0.243/11$ & $<5$ & $0.01$ \\
338$^{a}$& $2.28\pm^{0.17}_{0.11}$ 	& $197\pm^{1000}_{190}$	& $0.985/61$ 	& $<5$ & $0.45$ 
	& 	& 	&  	&  	&  \\
357$^{a}$& $2.22\pm^{0.24}_{0.18}$ 	& $1710\pm^{170}_{550}$	& $1.198/64$ 	& $<5$ & $0.16$ 
	& 	& 	&  	&  	&  \\\hline\hline
\end{tabular}

\tablefoot{
The table is divided into three parts: an additional black-body component was required to find
the best-fit model, the addition of partial covering and, finally cases in which the low did not allow us 
to reject any of the components. $KT$ is given in units of eV
and the intrinsic equivalent hydrogen column in units of $10^{20}$ atoms cm$^{-2}$. 
$^{**}$ denotes a parameter that XSPEC could not calculated, with an error bar as preceise as $90$ per cent confidence. \\
${\dagger}$ If the probability of the F-test is low (close to zero)  then it is reasonable to add the extra model component.\\
(a) The fit is insensitive to either the $N_{H,z}$ or $f$ parameters of the PCF model.\\
(b) The X-ray spectra could not fitted by the PCF model.\\
(*) The soft excess is extremely strong, invalidating the use of an abosrbed PL fitting over the whole X-ray spectra.\\
We could not analyse the X-ray spectra of the sources 15, 163, 324, 335, and 355 because of the small number of counts.
Furthermore, the sources 111, 161, 189, 198, and 302 were not analysed using any of these models, because their 
hard X-ray spectra is dominated by background, reducing the detectability of their source counts.}

\end{table*}

Table~\ref{tab:MissingAGN_fit_SE} shows the best-fit model parameters for each source.
The addition of either a BB or PCF component provides a good match to the observed spectra in almost all objects 
with more than 100 counts, and correspondingly provides a close fit according to the $\chi^2$ test, than the
absorbed-PL model (i.e. $\Delta\chi^2$ and/or $P_{F-test}$ 
in Table~~\ref{tab:MissingAGN_fit_SE}).  
An additional BB component was required to achieve good spectral fits in about one-third of these objects 
(first part of the Table~~\ref{tab:MissingAGN_fit_SE}); 
the inferred electron temperatures are found to be in the range of $100-200$ eV, which is slightly higher 
than those found for classic AGNs. Such high temperatures could be explained by the presence of a slim 
accretion disc in which the temperature is raised by photon trapping \citep{Abramowicz1988,Mineshige2000}. 
For another third of the missing-AGN subsample (second part of the Table~\ref{tab:MissingAGN_fit_SE}), 
the best-fit models were achieved by the addition of a PCF component. 
The spectral fits for a partial covering model indicate that there were variations in both the absorption column 
density $N_{H,z} =(1-6)\times 10^{21}$ cm$^{-2}$ and covering factor $f=0.6-0.9$. The measured strength
of the non-absorbed X-ray primary emission from the neutral material is $<1-f>=0.32$ ($\ge 0.2$ for the great 
majority) being slightly higher than those expected by type-2 AGNs ($\le 0.1-0.2$).

For the remaining third of the sources, the low quality of their X-ray spectra did not allow us to choose between
the various possible models. In summary, the X-ray spectra of the subsample of missing-AGNs were closely fitted by a rather steep
power-law, to which a soft excess apparent at energies $\lesssim 2$ keV should be added, when data of sufficiently high quality
becomes available. \emph{This is totally in line with the assumption that this population is largely dominated by NLS1s}.\\
 
\subsection{True-SF subsample}

In parallel, we conducted an X-ray spectral analysis of the true-SF subsample. We were able to perform the spectral 
analysis of $8$ of the $38$ sources only, because of the poor X-ray spectral quality of the remaining $30$.\\

The X-ray spectra of local SF galaxies in the $0.5-10$ keV band can be described by a combination of warm thermal
emission (with typically $kT\sim0.6-0.8$ keV) dominating at energies $\leq1$ keV, and a power-law spectrum responsible for 
producing the bulk of their \he~ keV flux. 
The latter component has various interpretations, either in terms of an extremely hot ($kT\geq5$ keV)
thermal component or a $\Gamma\sim2$ power-law model for high mass X-ray binaries. Given that SF
spectra often also exhibit strong collisionally excited emission lines, we used a MEKAL model \citep{Mewe1985,Mewe1986}
to fit the thermal component. This component appeared at soft X-ray energies (below $1-2$ keV), and we added a PL component
to fit the hard X-ray spectrum. The X-ray spectra were modelled by adding both component which resulted in good
fits for only 3 out of 8 objects (2XMM J140920.0+262219, 2XMM J093401.9+551428, and 2XMM J122254.6+154916, see 
Table~\ref{tab:MissingAGN})
with $kT\sim0.6-3$ keV and $\Gamma\sim2.2$; the resulting parameters were consistent with those expected for a SF galaxy.
The thermal component contributes significantly over the $0.3-10$ keV range, supplying $\sim20\%$ of the total flux. Hence,
a hot gas starburst component was found to be present in the spectra of these three objects. The improvement to the fits when a MEKAL
component was added to a PL in five of the remaining eight objects, was minimal with $\Delta\chi^2<5$ for two additional degrees 
of freedom (see Table~\ref{tab:SF_fit}). Hence, their X-ray spectra were modelled by a PL only, which resulted in an 
acceptable fit; the resulting parameters $\Gamma\sim1.7-2$ were compatible with those for 
a SF galaxy, as expected. We therefore restricted our analysis of the remaining $30$ sources 
to an inspection of the hardness ratios. In Figure~\ref{fig:HR2_vs_T_and_XO}, we can see that the 
hardness ratio of all sources covers a wide range of values, and therefore we are unable to reach any firm conclusion 
based on these data. However, the invariably low values of the HR for these particular sources are consistent 
with them being dominated by a thermal 
spectrum, in full agreement with the expectation for SF galaxies. \\

As a final additional test, we fitted the missing-AGN X-ray spectra as if they were true-SF sources. We found that the
soft X-ray excess is not properly described in terms of a thermal (MEKAL) emission in that case.


\begin{table*}[!htb]
        \caption[]{\emph{Star-forming galaxies}. Best fit models for the star forming galaxies.}
        \label{tab:SF_fit}

\centering
\begin{tabular}{lcccccc}\hline\hline
Source 	& Model	 	&  $\Gamma$ 		  & $KT$ 		   & $\chi^2/\nu$ & $F_{test}$ 	& $\Delta\chi^2$ \\ \hline

008	& A+B		& $2.16\pm^{0.51}_{0.41}$ & $0.901\pm^{0.28}_{0.12}$ & $8.80/8$ & $>0.97$ & $\sim13$ \\ 
079	& A+B$^{*}$    	& $2.35\pm^{0.16}_{0.16}$ & $3.71\pm^{0.88}_{3.14}$  & $103.25/98$ & $\sim 1$ & $>200$ \\ 
233	& A+B 	 	& $2.11\pm^{0.08}_{0.07}$ 	& $0.62\pm^{0.05}_{0.05}$ 	& $104.68/90$ & $\sim 1$&  $> 100$\\ 
\hline\hline
056	& A		& $1.84\pm^{0.22}_{0.20}$ &			     & $21.39/19$ & & \\ 
149	& A		& $1.83\pm^{0.42}_{0.38}$ & & $4.76/5$ & $0.0927$ & $<6$ \\ 
164	& A		& $1.94\pm^{0.31}_{0.29}$ & & $17.06/14$ & $0.2165$ & $<1$ \\ 
246	& A	 	& $1.88\pm^{0.21}_{0.19}$	& 				& $13.75/22$  & & \\ 
251	& A		& $1.40\pm^{0.27}_{0.25}$	&  				& $6.35/5$    & $0.63$ & $<4$ \\ \hline
\end{tabular}
\tablefoot{
A:{\tt power law}, B: {\tt mekal}
$^b$ Galactic neutral hydrogen column density (in units of $10^{20}$ atoms cm$^{-2}$,
from \citet{Dickey1990}.
$^{*}$ Intrinsec absoption with a $N_{h,int} = 2.1\pm^{0.2}_{0.3}$ in units of $10^{21}$ atoms cm$^{-2}$
}
\end{table*}


\begin{figure*}[!Hht]
        \centering
        \includegraphics[width=4.0cm,height=4.2cm,angle=-90]{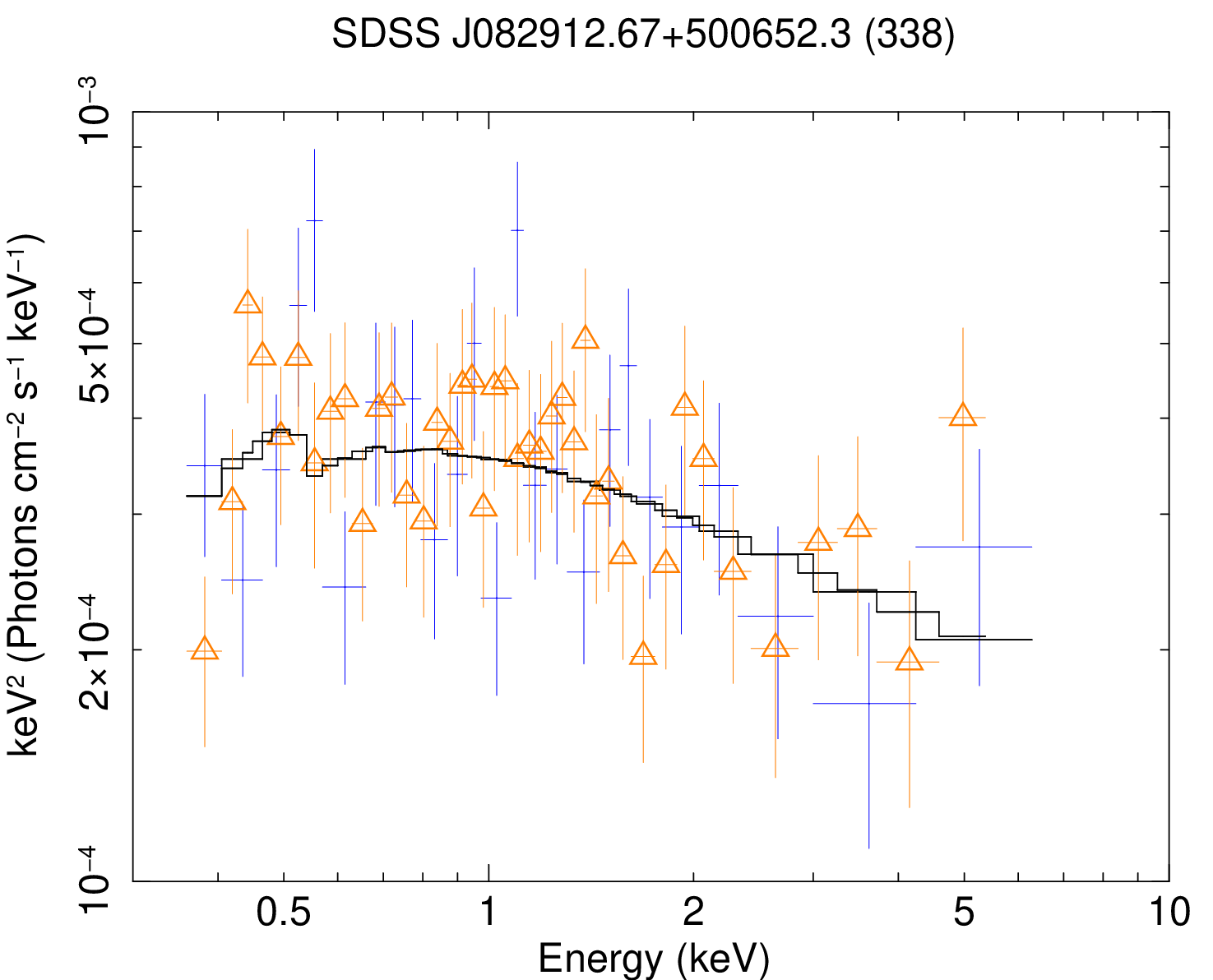}
        \includegraphics[width=4.0cm,height=4.2cm,angle=-90]{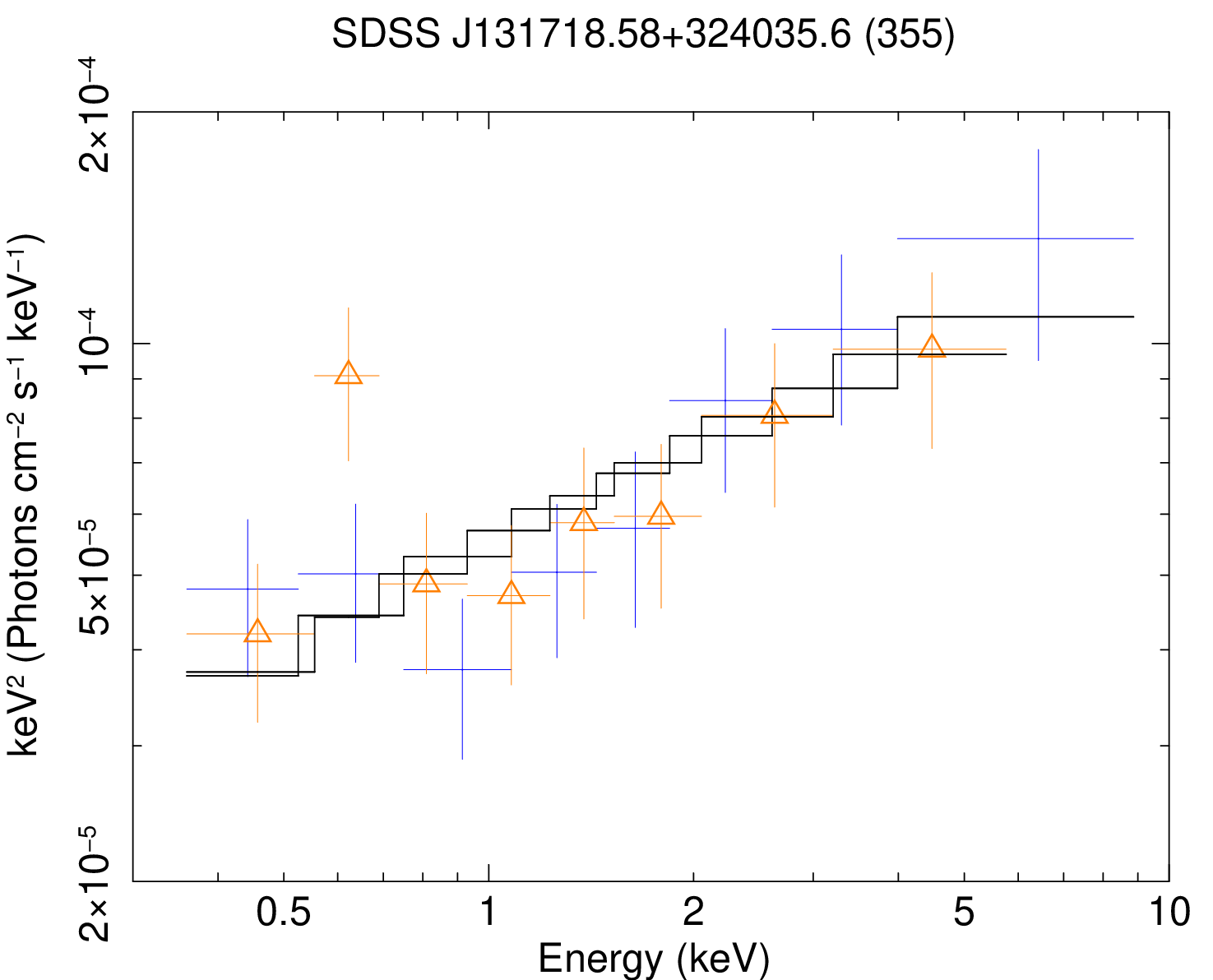}
        \includegraphics[width=4.0cm,height=4.2cm,angle=-90]{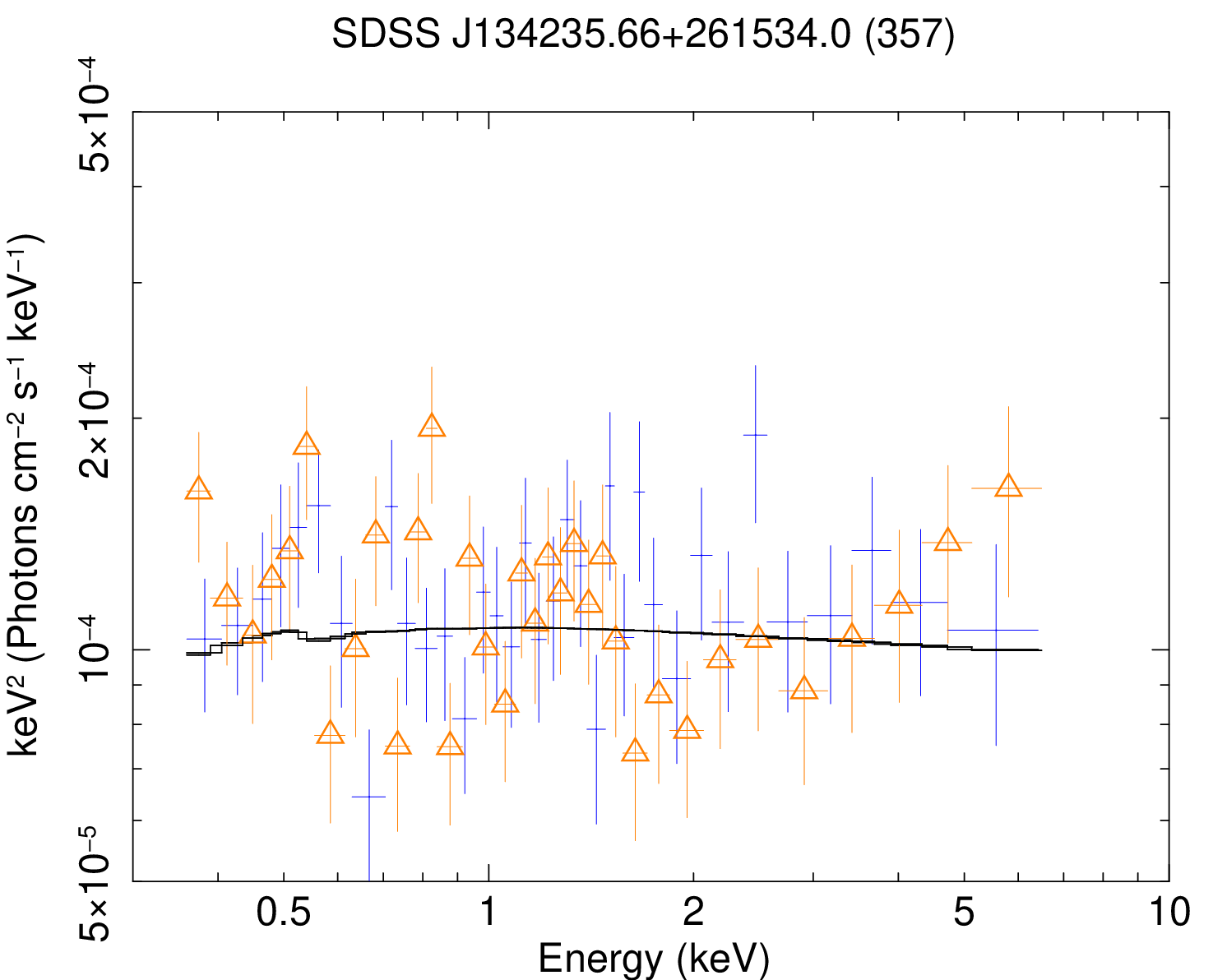}
        \begin{center}
                {\bf Figure \ref{fig:Spectra}.} \emph{Continued}
        \end{center}
\end{figure*}

\subsection{Missing-AGNs versus type-2 AGNs}
We then assessed the different nature of the missing-AGN sources in terms of spectral fitting of known type-2 AGNs. 
To avoid composite objects we adopted the \citet{Kewley2001} criterion to secure optically classified AGNs.
As for in the missing-AGN population, we only used the sources with a minimum of 50 counts in at least one detector. 
This requirement resulted in the selection of 56 AGNs. From inspection of the 
X-ray data, we excluded 15 of these objects that were dominated by the background above $\sim$~2-3 keV. 
Finally, we removed those sources that had previously been classified as LINERS. The final sample contains 34 bona-fide
type-2 AGN candidates. \\

\noindent
The results of our analysis are as follows:
\begin{enumerate}
    \item The spectra of $10$ type-2 AGN ($\sim29\%$) are best-fitted with a simple power-law. The average $\Gamma$ 
        obtained as a function of the $2-10$ keV flux is softer ($<\Gamma>=1.8$) than the typical
        values of $\sim 1.9$ and $\sim 2.1$ found in the unabsorbed AGN and the missing-AGN subsamples, respectively. 
    \item The spectra of $9$ AGNs ($\sim26\%$) could be most closely fitted by the addition of intrinsic absorption at the level
        of a few $\times 10^{22}$ cm$^{-2}$ and the average $\Gamma$ is also $<\Gamma>\cong1.8$.
    \item The addition of another PL component as a proxy for a partially covered absorber or scattering of 
        the AGN light, was required to achieve good spectral fits in another $14$ AGN ($\sim41\%$). We found that the
        average photon index is $<\Gamma>=2.2$ and the average intrinsic column denssity $<N_{H,z}>=5\times10^{23}$
        cm$^{-2}$. The measured amount of X-ray absorption for the missing-AGN subsample are lower (by two orders of magnitude),
        falling in the low part of the column density distribution of type-2 AGNs. The average measured strength of 
        the non-absorbed X-ray primary emission by the neutral material 
        was measured to be softer ($\sim 8\%$) than the average of the Missing-AGN subsample. 
         Soft excess emission is not detected with an F-test significance $>99\%$ in the vast majority of the objects.
    \item Finally, we found a significant X-ray soft excess in only $1$ AGN ($\sim3\%$), which is clearly negligible 
         compared with the strong soft excess displayed by the missing-AGN subsample.
\end{enumerate}

The conclusion is therefore that the X-ray properties of the missing-AGN population, which we interpret as being dominated by NLS1s,
clearly differ from those of the type-2 AGNs in our sample.

\section{Discussion}
\label{section4}

We investigated the nature of galaxies that are optically
classified as star-forming, but whose X-ray luminosities are
in excess of $10^{42}$ \lum, hence indicative of an AGN
(missing-AGN subsample) by exploring both the X-ray
and optical properties of a sample of NELGs. The availability
of catalogues at different wavebands with wide sky coverage,
allowed us to assemble optical (SDSS-DR7) and X-ray (2XMMi
catalogue) information for a large sample of NELGs.
A total of $1729$ NELGs fall in the region covered by \xmm 
observations, with which the 2XMMi catalogue was built.
Out of these we find $211$ X-ray detections, and $1518$ upper limits.

For the $211$ NELGs detected in X-rays, we compared 
the optical classification based on the BPT
diagram with their hard X-ray luminosity used as an indicator
of AGN activity. We found that about 40\% of the
BPT-AGN subsample exhibit low luminosities not exceeding
$10^{42}$ \lum~ (Weak-AGN subsample). Using other optical
spectral features, such as the [OI] and [OII] emission lines,
we find that 13\% of these Weak-AGN subsample are classical
LINERs and 29\% are likely to be weak-[OI] LINERs. The
LINER X-ray spectral energy distribution can be interpreted as
a combination of a soft thermal component plus a hard power
law \citep[see ][]{Gonzalez2006}. This soft component
may arise from circumnuclear star formation that could also
explain their emission-line ratios.\\

However, the most striking result of this work is that virtually all
sources in the missing-AGN subsample which have been
classified as BPT-SF according to the BPT diagram and yet
display clear signs of AGN activity in the X-ray band are
unquestionably NLS1. To investigate this, we have calculated
the hard X-ray luminosity, thickness parameter, X-ray-to-optical
flux ratio and hardness ratio for the whole NELG sample. We
found that it is not clear that we can distinguish between SF and
AGN galaxies using a single criterion (Kauf03 or \lx). However,
the combined use of $X/O$ flux ratio, $T$ and $HR$ allows us to
distinguish between SF galaxies and AGNs. For our sample of
NELGs including objects of high \lx and low redshift
($z < 0.4$), we found that the distributions of values of both
the thickness parameter ($T$) and X-ray-to-optical flux ratio ($X/O$)
are bimodal, with the two populations being separated by about $T\sim1$
and $X/O\sim0.1$, respectively. We noted dichotomies in the
BPT-SF population, i.e. between the missing-AGN subsample
and True-SF galaxies, and on the other hand, in the BPT-AGN
subsample, i.e. between the weak-AGN and strong-AGN subsamples.\\

We performed several tests to determine whether the
emission coming from missing-AGNs arises as a result of
star formation processes, or as AGN activity. The dichotomy
in the BPT-SF population is directly linked to the values of
\Hb~ FWHM: all the galaxies with high \lx~ exhibit the broadest
widths in the \Hb~ line, from $\sim 600$ to $1200$ km/s, whilst the
remainding sources with \lx$<10^{42}$ \lum, display an \Hb~ FWHM
$\lesssim600$ km/s. Indeed, the missing-AGN subsample has
high values of both $X/O$ and $T$ and low measured hardness 
ratio for each source. So we conclude that these missing-AGNs
are NLS1 candidates whilst the rest (true-SF population) are
consistent with being SF galaxies. \\

Strong supporting evidence for the NLS1 nature of the missing-
AGN population comes from a spectral analysis of their X-ray
emission. The X-ray spectral properties of the missing-AGN
subsample appear to be quite uniform. These spectra are well
reproduced by the combination of black body or a partial covering
absorption on a steep power-law component at hard X-ray
energies. A complete study of their spectral energy distribution
is in progress by Castell\'o-Mor et al. (in prep.). Furthermore, we
have established evidence of a missing-AGN population displaying 
a soft X-ray excess, whenever spectra of sufficient $S/N$ is available to model them. 
The missing-AGN population have \Hb~ lines FWHMs
larger than $600$ km/s, and often display strong FeII emission.\\

Therefore, we conclude that the population of the missing-AGN
subsample is entirely constituted of NLS1s
with very moderate broad-line velocities ($600$ km/s $\lesssim$ FWHM(\Hb)
$\lesssim 1200$ km/s), which have $X/O>0.1$, $T>1$ and the vast majority
have strong soft X-ray excess.

\begin{acknowledgements}
We are grateful to the referee for comments that helped improve
the paper.
N. Castell\'o-Mor gratefully acknowledges the warm hospitality in 2011 of the Physics Department 
of the University of Durham where part of this work was done, and the Spanish Ministerio de Ciencia e 
Innovaci\'on for a pre-doctoral fellowship. 
NC-M, XB, LB, and FJC  acknowledge partial financial support from the Spanish Ministerio 
de Ciencia e Innovaci\'on, through project AYA2010-21490-C02-01. 
\end{acknowledgements}

\bibliographystyle{aa}          


\end{document}